\documentclass[structabstract]{aa}  
\usepackage{longtable}
\usepackage{supertabular}
\usepackage{graphicx}
\usepackage{natbib}
\usepackage{txfonts}
\bibpunct{(}{)}{;}{a}{}{,}
\begin{document}
   \title{Trigonometric parallaxes of 6.7 GHz methanol masers}

   \author{K.~L.~J. Rygl\inst{1}\fnmsep\thanks{Member of the International Max Planck Research School
(IMPRS) for Astronomy and Astrophysics at the Universities
of Bonn and Cologne},
          ~A. Brunthaler\inst{1},~M.~J. Reid\inst{2},~K.~M. Menten\inst{1},~H.~J. van Langevelde\inst{3,4}, \and Y. Xu\inst{5}
%\fnmsep\thanks{Just to show the usage  of the elements in the author field}
          }

\institute{Max-Planck-Institut f\"ur Radioastronomie (MPIfR),
              Auf dem H\"ugel 69, 53121 Bonn, Germany\\
              \email{[kazi;brunthal;kmenten]@mpifr-bonn.mpg.de}
             \and
             Harvard-Smithsonian Center for Astrophysics, 60 Garden Street,
  Cambridge, MA 02138, USA\\
             \email{reid@cfa.harvard.edu}
             \and Joint Institute for VLBI in Europe, Postbus 2, 7990 AA Dwingeloo, the Netherlands\\
\email{langevelde@jive.nl}
\and Sterrewacht Leiden, Leiden University, Postbus 9513, 2300 RA Leiden, the Netherlands
\and Purple Mountain Observatory, Chinese Academy of Sciences, Nanjing 21008, China\\
\email{xuye@pmo.ac.cn}
             }

%  \date{Received\Accepted }

\authorrunning {K.~L.~J.~Rygl et al.}
\titlerunning {Trigonometric parallaxes of 6.7 GHz methanol masers} 
% \abstract{}{}{}{}{} 
% 5 {} token are mandatory
 
  \abstract
   {}
   { Emission from the 6.7 GHz methanol maser transition is very strong, is relatively stable, has small internal motions, and
     is observed toward numerous massive star-forming regions in the
     Galaxy. Our goal is to perform high-precision astrometry using this maser
     transition to obtain accurate distances to their host regions.} 
   {Eight strong masers were observed during five epochs of VLBI observations
     with the European VLBI Network between 2006 June, and 2008 March.}
   {We report trigonometric parallaxes for five star-forming regions, with accuracies as good as $\sim22~\mathrm{\mu}$as.  Distances to these
     sources are $2.57^{+0.34}_{-0.27}$~kpc for ON\,1, $0.776^{+0.104}_{-0.083}$ kpc for L\,1206, $0.929^{+0.034}_{-0.033}$ kpc for L\,1287, $2.38^{+0.13}_{-0.12}$ kpc for NGC\,281-W, and $1.59^{+0.07}_{-0.06}$ kpc for S\,255.
     The distances and proper motions yield the full space motions of the star-forming regions hosting the masers, and we find that 
     these regions lag circular rotation on average by $\sim$17~km~s$^{-1}$, a
     value comparable to those found recently by similar studies.
     }
   {}


   \maketitle
%
%________________________________________________________________

\section{Introduction}

Accurate distances and proper motions are crucial for studies of the structure 
and kinematics of the Milky Way. Since we are in the Galactic plane, it is not
easy to determine the spiral structure of our Milky Way \citep{reid:2009b}. 
Massive star-forming regions trace the spiral arms and are objects well-suited to revealing the structure of the Milky Way.

Determining the fundamental physical properties of
individual objects, such as size scales, masses, luminosities, and ages, also depends critically on distance. For example, the distance to the Orion
Nebula, determined by a number of trigonometric parallax measurements using
radio continuum emission from stars in its associated cluster \citep{menten:2007,Sandstrom:2007}, water masers \citep{hirota:2007}, and SiO masers \citep{kim:2008}, turned out to be 10\% less than previously
assumed, resulting in 10\% lower masses, 20\% fainter luminosities, and 20--30\% younger ages for the stars in the cluster.

The most fundamental and unbiased method of measuring distances is the
trigonometric parallax, which depends only on geometry and is therefore free
of any astrophysical assumptions. To achieve the sub-milliarcsecond astrometric accuracy at optical
wavelengths requires space-borne observations. The first
dedicated optical satellite for this purpose was ESA's Hipparcos
\citep{perryman:1995}. With its accuracy of 0.8--2\,mas, it was capable of
measuring distances up to $\sim$200\,pc -- only a small fraction of the Milky
Way. A new optical astrometry satellite GAIA, to be launched in 2012, will be
two orders of magnitude more accurate and push the optical parallaxes to
Galactic size scales \citep[e.g.,][]{lindgren:2009}. Despite the high accuracy, GAIA
will suffer from dust extinction in the Galactic plane, in particular in the
spiral arms and toward the Galactic center. Here radio astronomy can provide
an important complement to GAIA, because radio wavelengths do not suffer from dust
extinction, and Very-Long-Baseline Interferometry (VLBI) phase-referencing
techniques can provide accuracies better than 10\,$\mathrm{\mu}$as, allowing
accurate distances with errors less than 10\% out to 10\,kpc
{\cite[see for example][]{honma:2007,hachisuka:2009,reid:2009a}.     

Strong and compact radio sources, such as molecular masers, are ideal targets
for trigonometric parallax measurements. Masers are frequently found in (dusty)
star formation regions (SFRs) and asymptotic giant branch stars.
The most numerous interstellar masers are the 22\,GHz water masers, followed in number by the 6.7\,GHz
methanol masers; to date several hundreds of these masers have been
found across the Galaxy \citep[e.g.,][]{pestalozzi:2005, green:2009}. They are exclusively
associated with early stages of massive star formation, observed both prior to
the development of an ultra compact H{\sc ii} (UCH{\sc ii}) region and
coexistent with these \citep{menten:1991, ellingsen:2006, pestalozzi:2007}. 
Maser modeling indicates that the 6.7\,GHz emission is likely to be excited via
radiative pumping by warm dust heated by newly formed high-mass stars \citep{sobolev:2007}. 

Sources of methanol masers have long lifetimes of $\sim$\,$10^4$\,year \citep{walt:2005}. 
Their velocity spread is typically within $\pm5$\,km~s$^{-1}$ about the systematic
velocity of the molecular core, and single maser features are usually narrow
($\sim$1\,km~s$^{-1}$). The kinematics suggest that methanol masers originate
from different regions than water masers, which have wider velocity spreads and
are produced in protostellar outflows \citep{menten:1996}. To summarize, the strength,
ubiquity, long lifetimes, and small internal motions make the 6.7\,GHz methanol
maser very suitable for astrometric purposes. 

Here we present results from observations with 
the European VLBI Network (EVN) of eight
strong methanol masers belonging to well-known massive SFRs in the outer part
of the Milky Way: ON\,1, L\,1287, L\,1206, NGC\,281-W, MonR\,2, S\,252, S\,255 and S\,269. 
Recently, H$_2$O maser or 12.2\,GHz methanol maser parallax measurements have been reported for three of these regions
(NGC\,281-W, \citeauthor{sato:2008} \citeyear{sato:2008}; S\,252, \citeauthor{reid:2009a} \citeyear{reid:2009a}; S\,269, \citeauthor{honma:2007} \citeyear{honma:2007}), allowing a valuable cross check with our measurements. The results of
this present work are crucial for star formation studies in these regions,
and, together with a larger sample of parallaxes, will help to understand the
structure of the Local and Perseus arms. 

In this paper we report on the first parallax measurements made
with the EVN. Results presented here replace preliminary, less
accurate, and less complete measurements \citep{rygl:2008}.
In the next section we describe the observations and data analysis, and in Sect.~3 we explain the method and fitting of the parallax and proper motion. The results are then presented per source in Sect.~4 and discussed in
Sect.~5. The conclusions are summarized in Sect.~6. 
%__________________________________________________________________

\section{Observations and data analysis}

\begin{table}
\begin{minipage}[t]{\columnwidth}
\centering
\renewcommand{\footnoterule}{}
\caption{\label{ta:obs} Observation information}
\begin{tabular}{l c l}
\hline\hline
Program & Date & Participating Antennas\footnote{The full names of the antennas are: Jb -- Jodrell Bank; Wb --  Westerbork (in single dish mode); Ef -- Effelsberg; Mc -- Medicina; Tr -- Torun; Nt -- Noto; Hh -- Hartebeesthoek; On -- Onsala; Da -- Darnhall; Cm -- Cambridge; Y1 -- single EVLA antenna.}\\
\noalign{\smallskip}
\hline
\noalign{\smallskip}
%        &      & Jb & Wb & Ef & Mc & Tr & Nt & Hh & On & Da & Cm & Yy\\
EB32A   & 10 Jun 2006 & Jb~Wb~Ef~Mc~Tr~Nt~Hh~On~Da\\
%$\surd$&$\surd$&$\surd$&$\surd$&$\surd$&$\surd$ &$\surd$&$\surd$ &$\surd$&...&... \\ 
EB32B   & 18 Mar 2007 &Jb~Wb~Ef~Mc~Tr~Nt~Hh~On~Cm\\%$\surd$&$\surd$&$\surd$&$\surd$&$\surd$&$\surd$ &$\surd$&$\surd$ & ...   &$\surd$&... \\
EB32C   & 17 Jun 2007 &Jb~Wb~Ef~Mc~Tr~Nt~Hh~On~Cm\\%$\surd$&$\surd$&$\surd$&$\surd$&$\surd$&$\surd$ &$\surd$&$\surd$ & ...&$\surd$&... \\
GB63A   & 31 Oct 2007 &Jb~Wb~Ef~Mc~Tr~Nt~Hh~On~Y1\\%$\surd$&$\surd$&$\surd$&$\surd$&$\surd$&$\surd$ &$\surd$&$\surd$ &...&...&$\surd$ \\
GB63B   & 16 Mar 2008 &Jb~Wb~Ef~Mc~Tr~Nt~Hh~On~Y1\\%$\surd$&$\surd$&$\surd$&$\surd$&$\surd$&$\surd$ &$\surd$&$\surd$ &...&...&$\surd$ \\
\noalign{\smallskip}
\hline
\noalign{\smallskip}
\end{tabular}
\end{minipage}
\end{table}

The observations were performed with the EVN at five epochs between 2006 June,
and 2008 March. The exact dates of the observations and the participating antennas are listed in Table \ref{ta:obs}.
Each observation lasted 24 hours
and made use of \emph{ geodetic-like} observations to calibrate the tropospheric
zenith delays at each antenna \citep[see][for a detailed
discussion]{reid:2004, brunthaler:2005, reid:2009a}. A typical 
observing run consisted of four 6-hour blocks containing $\sim$1 hour of
geodetic-like observations, $\sim$10\,minutes of observation of fringe finders. The
remaining time was spent on maser/background source phase-referencing
observations. During each run, the average on-source time per maser was between $\sim$0.9 and
$\sim$1.2\,hours depending on the sky position, because half of our targets were low
declination sources that had a limited visibility.  

Using the technique of phase referencing, each maser was observed in a cycle
with two (or one) nearby ($\sim$$1\degr$--$2\degr$ separation) compact radio
quasars, which were used as background sources. 
The sources were switched every 2 minutes. Before our EVN observations, we had used the NRAO Very Large Array under project AB1207 in A-configuration to observe several compact NVSS \citep{condon:1998} sources within 1\degr\ from the maser source at two frequencies (8.4 and 4.8\,GHz) to check their spectral index and compactness. 
For the best candidates, we determined their position to a sub-arcsecond accuracy to use them as a position reference in the EVN observations. 
Additionally, several known VLBA calibrators were used, namely: J2003+3034 and J0613+1708 \citep{fomalont:2003}, J0603+215S(9) and J0613+1306 \citep{ma:1998}, J0047+5657 \citep{beasley:2002}, J2223+6249 \citep{petrov:2005}, J0035+6130 \citep{petrov:2006}, and J0606-0724 \citep{kovalev:2007}.
Table
 \ref{ta:sour} lists the positions of the masers and their background sources.

The observations
were performed with eight intermediate frequency bands (IF) of 8\,MHz width, each in dual circular
polarization sampled at the Nyquist rate with 2 bits per sample, yielding a recording rate of 512\,Mbps. The data were correlated
in two passes at the Joint Institute for VLBI in Europe (JIVE), using an
integration time of 0.5\,seconds, affording a field of view of $1\rlap{$.$}\,'2$ 
(limited by time-averaging smearing). The maser data were correlated using one 8\,MHz IF band with
1024 spectral channels, resulting in a channel separation of 7.81\,kHz
or 0.41\,$\mathrm{km~s}^{-1}$ at 6.7\,GHz. The quasar
sources were correlated in continuum mode with eight IFs of 8\,MHz width
with a channel separation of 0.25\,MHz.   

The data were reduced using the NRAO Astronomical Image Processing System
(AIPS). The \emph{geodetic-like} observations were reduced separately, and
tropospheric delays were estimated for each antenna. The
data were reduced following the EVN guidelines, applying parallactic angle and
ionospheric delay corrections. The JIVE correlator model uses Earth's
orientation parameters, which are interpolated from the appropriate daily-tabulated values,
so it is not necessary to correct them after the correlation. 
The ionospheric delays were based on the JPL GPS--IONEX total vertical electron
content maps of the atmosphere. Amplitudes were
calibrated using system-temperature measurements and standard gain curves. 
A ``manual phase-calibration'' was performed to remove delay and phase
differences between the IFs. The Earth rotation was corrected for with the task `CVEL'. 
For each maser, a spectral channel with one bright and compact maser spot was
used as the phase reference. The data was Hanning-smoothed to minmize Gibbs
ringing in the spectral line data. To avoid the strong fluctuations caused by the
bandpass edges, the outer two channels in each IF were discarded
\citep[following ][]{reid:2009a}. The positions of the masers
and background sources were extracted by fitting 2D Gaussians to the maps. 

Here, we report the results on the first five sources for which we have
completed the analysis: ON\,1, L\,1206, L\,1287, NGC\,281-W, and S\,255. For the other
sources, S\,252, S\,269, and MonR\,2, we had problems with the calibration of the
data, which were likely caused by residual atmospheric delay, so that we were unable to achieve sufficient accuracy for a parallax
measurement.

\begin{table*}
\begin{minipage}[t]{\textwidth}
\centering
\renewcommand{\footnoterule}{}
\caption{\label{ta:sour} Source information}
\begin{tabular}{l c c r r l l}
\hline\hline
Source  & R.A. (J2000) & Dec. (J2000) & $\phi$ \footnote{Separation, $\phi$,
  and the Position Angle (east of north), P.A.~, between the maser and the background source.}& P.A.$^a$ & Brightness\footnote{ The brightness and restoring beam (east of north) are listed for the second epoch.} & Restoring Beam$^b$\\
& (h:m:s) & ($^\circ$ : ' : '')& (\degr) & (\degr) & ($\mathrm{Jy~beam^{-1}}$) & (mas, mas, deg)\\
\noalign{\smallskip}
\hline
\noalign{\smallskip}
ON\,1 ...............         & 20:10:09.074  & +31:31:35.946 & ... & ...&1--7&$5.7\times3.9,$$~~86$\\
J2003+3034 ...  & 20:03:30.244  & +30:34:30.789 & 1.71 &--124&0.124&$5.3\times3.7,$$~~76$\\
J2009+3049 ...   & 20:09:17.588  & +30:49:24.580 & 0.73 &--14&0.008&$5.6\times3.7,$$~~88$\\
%&&&&\\
\noalign{\smallskip}
L\,1206 .............       & 22:28:51.407 & +64:13:41.314  & ...  & ... &0.3--1.4 & $5.7\times4.0,$$-80$\\
J2223+6249 ...  & 22:23:18.097 & +62:49:33.805  & 1.53 &--157 & 0.093&$5.7\times3.9,$$-84$\\
J2225+6411 ...   & 22:25:27.993 & +64:11:15.030  & 0.37 &--96 &0.002 &$5.6\times3.8,$$-80$\\
\noalign{\smallskip}
L\,1287 .............  & 00:36:47.353 & +63:29:02.162& ...  & ...&0.5--4.1&$6.0\times4.1,$$-57$\\
J0035+6130 ... & 00:35:25.311 & +61:30:30.761& 1.98 &--176&0.092&$6.0\times4.1,$$-58$\\
J0037+6236 ...  & 00:37:04.332 & +62:36:33.310& 0.88 &178&0.061&$5.9\times4.2,$$-57$\\
\noalign{\smallskip}
NGC\,281-W ... & 00:52:24.196& +56:33:43.175 & ... & ... &1--12&$5.7\times4.2,$$-56$\\
J0047+5657 ...  & 00:47:00.429 & +56:57:42.395 & 0.84 & 62&0.142&$5.6\times4.0,$$-65$\\
J0052+5703 ... & 00:52:54.303 & +57:03:31.460 & 0.50 & --8&0.010&$5.6\times3.9,$$-62$\\
\noalign{\smallskip}
MonR\,2 ........      & 06:07:47.862 & -06:22:56.518 & ... & ...&9--22&$8.6\times4.4,$$~~42$\\
J0606-0724 ... & 06:06:43.546 & -07:24:30.232 & 1.06 & --165&0.135&$9.0\times4.1,$$~~40$\\
\noalign{\smallskip}
S\,252 ..............       & 06:08:53.344 & +21:38:29.158 &... & ...&17-44&$6.4\times3.9,$$~~56$\\
J0603+215S .. & 06:03:51.557 & +21:59:37.698 & 1.22& --73 &0.019&$6.7\times3.8,$$~~54$\\
\noalign{\smallskip}
S\,255 ..............       & 06:12:54.020 & +17:59:23.316 & ... & ...&4&$6.9\times3.9,$$~~56$\\
J0613+1708 ... & 06:13:36.360 & +17:08:24.946 & 0.87 & 169 &0.040&$7.1\times3.9,$$~~46$\\
\noalign{\smallskip}
S\,269 ..............       & 06:14:37.055 & +13:49:36.156 & ... & ... &0.2--1.1&$6.8\times3.9,$$~~58$\\
J0613+1306 ... & 06:13:57.693 & +13:06:45.401 & 0.73 & --167&0.087&$6.8\times3.8,$$~~57$ \\
\noalign{\smallskip}
\hline
\end{tabular}
\end{minipage}
\end{table*}

\section{Method and fitting}

The absolute positions of the masers were not known with milli-arcsecond
accuracy before our first observation. Only after the first epoch was
analyzed were accurate positions determined. As a result, different correlator
positions were used between the first and following epochs. 
In the first epoch, position errors above $0\rlap{$.$}\,''2$ made it difficult to
calibrate the phases because of a high fringe rate, which could not be interpolated
correctly when applied to the data. Shifting the maser into the phase center turned out to be very time-consuming with available software because AIPS and the EVN correlator use apparent
positions calculated at different times (AIPS at 0 UT of the first observing
day versus the EVN correlator at the end of the first scan).  Even an infinitesimally small
shift in the epoch coordinate using the AIPS tasks `CLCOR' or `UVFIX' can result in a substantial position change and phase shifts from the old EVN apparent position to the new AIPS apparent position at 0 UT. 
In a parallax experiment, accurate and consistent
positions throughout the experiment are crucial. Since it was difficult to calibrate the phases for masers with a large
position offset or to shift these to the phase center, several sources in the first epoch
could not be used for phase referencing. Although it should be possible to correct for this difference in registration, it would have
required a software effort beyond the scope of the current project. Therefore
we chose to discard some of the first-epoch data.

Most of the ionospheric delay is removed using measurements of the total
electron content of the ionosphere obtained from dual-band global positioning system
measurements \citep[e.g.,][]{ros:2000}. The expected residual delay is still a few centimeters, comparable to
the residual tropospheric delay at the observing frequency. Our single frequency geodetic-like
observations do not allow a separation of the ionospheric and tropospheric contributions. The measured group-delays from the geodetic blocks are
interpreted as having a tropospheric origin, and the data are phase-delay-corrected
accordingly. An ionospheric group delay makes a different contribution to the
phase delay, so that the tropospheric correction can even deteriorate the
calibration when the group delay has a partially ionospheric origin.
Since the residual tropospheric delay is a few centimeters at 6 GHz and can easily be confused with an ionospheric residual, the tropospheric correction
is not expected to be very useful in improving the quality of calibration. 
As a test, all data were reduced with and without a 
``tropospheric'' delay correction.  In slightly more than half of the cases, the signal-to-noise ratio improved by applying the tropospheric correction. We noted a trend toward improvement with decreasing declination:  both high declination sources,
$\delta\simeq64\degr$, L\,1206 and L\,1287 improved in 2 out of 5 epochs; NGC\,281-W
at $\delta\simeq57\degr$ improved in 3 out 5 epochs; and ON\,1 at
$\delta\simeq31\degr$ even improved in 4 out of 5 epochs. 
Ionospheric delay saturates at zenith angles greater than $\sim$$70\degr$. However, the
tropospheric delay continues to grow rapidly at larger zenith angles and can 
dominate the ionospheric delay at zenith angles of $\sim80\degr$ \citep[see][]{thompson:1991}.

For each SFR, we found the emission to arise from a number of separate
maser spots. Most of the line profiles stretched over
several channels. Both spatially and/or spectrally different maser spots were
considered as distinct maser components. We inspected the behavior of the proper
motion for each maser spot relative to the reference spot. Maser spots with 
strong nonlinear proper motions or a large scatter of position about 
a linear fit were discarded. Only compact maser spots with well-behaved proper 
motions were used for the parallax fitting. 

The average internal proper motion of the maser spots ranged between 0.06 and
$0.23\,\mathrm{mas~yr^{-1}}$. Considering the distance of each maser, these proper
motions correspond to 0.5--1\,$\mathrm{km~s}^{-1}$, much lower than the
internal proper motions of water masers, which can reach up to
20--200\,$\mathrm{km~s}^{-1}$ \citep{hachi:2006}. The only exception was ON\,1,
which separates into two distinct maser groups with a relative proper motion of $0.52\,\mathrm{mas~yr^{-1}}$, or 6.3\,$\mathrm{km~s}^{-1}$. This particularity is
discussed in Sect.~3.1.

The parallaxes and proper motions were determined from the change in the
positions of the maser spot(s) relative to the background source(s). The data
were fitted with a parallax and a linear proper motion. Since the formal
position errors are only based on the signal-to-noise ratios determined from the images,
they do not include possible systematic errors from residual zenith delay
errors or source structure changes. This leads to a high reduced $\chi^2$
value for the fits, so we added error floors in quadrature to 
the positions until reduced $\chi^2$ values close to unity were reached
for each coordinate. 

First, we performed parallax and proper motion fits for each maser spot
relative to one background source. Then we made combined fits with respect to each background source, assuming one parallax but different proper motions for each maser
spot. Finally, we repeated this combined fit for both the background sources together.
The position measurements of different maser spots are not
independent, since systematic errors, such as an unmodeled atmospheric
delay, will affect all maser spots in a similar way. If these systematic
errors dominate, this will lead to unrealistically small errors.
The most conservative approach, which we adopted, is to assume that the systematic errors are
100\% correlated. Then the error of the combined fit has to be multiplied by
$\sqrt N$, where $N$ is the number of maser spots.    

However, this will overestimate the error, if significant random errors are
present (e.g., owing to maser blending and structural changes over time),
since the latter are not correlated between different maser spots.
Random errors can be reduced by averaging the positions of the different maser
spots \citep[following the approach of][]{bart:2008, hachisuka:2009}. We
calculated the average positions with respect to each background source after
removing their position offsets and proper motions. Then, we performed a
parallax fit on these averaged data sets relative to each individual
background source, and on both the background sources combined. This approach has the advantage that
we can reduce the random errors, while leaving the systematic errors intact. In
Table \ref{ta:data} we list the individual parallax and proper
motion fits to each maser spot, the combined fit of all the maser spots and the fit of the averaged data sets.  

For most masers, we observed two background sources. Some of these
background source pairs show a variation in their separation (mimicking a
proper motion) of up to 1--2\,mas~yr$^{-1}$ (on
average this apparent movement was $0.54\,\mathrm{mas~yr^{-1}}$). This
apparent movement is much greater than expected for extragalactic sources measured at 12\,GHz,
$<0.1\,\mathrm{mas~yr^{-1}}$ \citep{reid:2009a} or 22\,GHz,
$<0.02\,\mathrm{mas~yr^{-1}}$ \citep{brunthaler:2007}.  While at higher
frequencies, the radio emission is typically dominated by the flat spectrum
radio cores, the steep spectrum emission from the radio jets becomes stronger at lower frequencies. Thus, structure changes in these jets can lead
to apparent motions, even if these jets are unresolved. At the distances of the
respective masers, these apparent movements correspond to 2--10\,km~s$^{-1}$. 
Since we do not know which of the two background sources is responsible for
this movement, the apparent movements of the background source pairs were added in quadrature to
the final errors of the averaged proper motions of the masers. A summary of our parallaxes and proper motions is given in Table \ref{ta:sum}.

\begin{table}[!htb]
\begin{minipage}[t]{\columnwidth}
\centering
\caption{\label{ta:sum}Parallax and proper motion results}
\renewcommand{\footnoterule}{}
\begin{tabular}{l@{\,}r@{\,\,}r@{\,\,}r@{\,\,}r}
\hline\hline
Source & \multicolumn{1}{c}{$\pi$}&\multicolumn{1}{c}{$D_\pi$}
&\multicolumn{1}{c}{$\mu_\alpha$\footnote{ The proper motion,
      $\mu_\alpha$, includes the factor $\cos(\delta)$.}\fnmsep\footnote{The
    errors of the proper motion take the uncertainty resulting from an apparent nonzero movement between the background sources into account.}} & \multicolumn{1}{c}{$\mu_\delta\,^b$}  \\
       & \multicolumn{1}{c}{(mas)} & \multicolumn{1}{c}{(kpc)}       & \multicolumn{1}{c}{($\mathrm{mas~yr^{-1}}$)}&  \multicolumn{1}{c}{($\mathrm{mas~yr^{-1}}$)}\\%& (km~s$^{-1}$)\\
\noalign{\smallskip}
\hline
\noalign{\smallskip}
ON\,1 &$0.389\pm0.045$&$2.57^{+0.34}_{-0.27}$ &
$-3.24\pm0.89^c$&$-5.42\pm0.46$\footnote{For the proper motion of ON\,1, we took an average of the north and south components.} \\%$5\pm2.5$\\
\noalign{\smallskip}
L\,1206 &$1.289\pm0.153$&$0.776^{+0.104}_{-0.083}$ &$0.27\pm0.23$&$-1.40\pm1.95~$\\%$-12.1\pm1.2$\\
\noalign{\smallskip}
L\,1287 &$1.077\pm0.039$&$0.929^{+0.034}_{-0.033}$ & $-0.86\pm0.11$&$-2.29\pm0.56~$\\%$-24.8\pm2.3$ \\
\noalign{\smallskip}
NGC\,281-W &$0.421\pm0.022$&$2.38^{+0.13}_{-0.12}$ & $-2.69\pm0.16$ & $-1.77\pm0.11~$\\%-29.2\pm1.1&\\
\noalign{\smallskip}
S\,255 &$0.628\pm0.027$&$1.59^{+0.07}_{-0.06}$&$-0.14\pm0.54$&$-0.84\pm1.76~$\\%&$4.6\pm2.0$\\
\noalign{\smallskip}
\hline
\end{tabular}
\end{minipage}
\end{table}

\section{Individual sources}
\subsection{Onsala~1}

\begin{figure*}[!htb] 
\centering
\includegraphics[width=\textwidth]{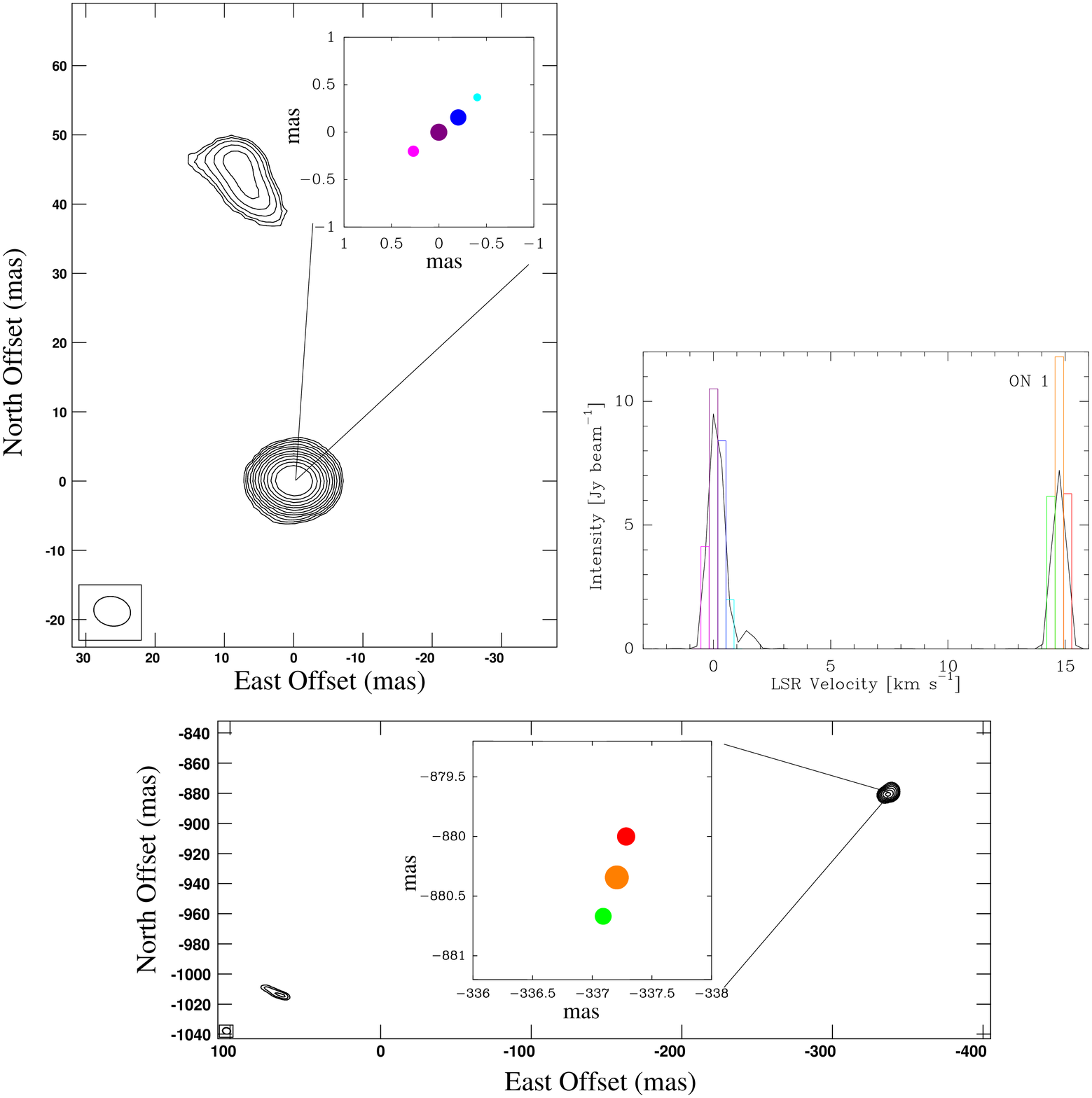}
\caption{\label{fig:on1} Velocity-integrated maps of the northern (top, left)
  and southern (bottom) maser groups in ON\,1, together with the spectrum (top,
  right). Position offset (0,0) corresponds to the position listed in Table \ref{ta:sour}. Maser spots are indicated with color codes for different radial
    velocities. The areas of the colored circles and the colored histogram entries in the spectrum are scaled to the peak flux of that spot. The black line in the spectrum is the intensity of the maser within a selected surface, which is not necessarily the same as the intensity of the maser spot retrieved from a Gaussian fit. For the northern group, the contour levels start at 0.1
  Jy$\mathrm{~beam^{-1}~km~s^{-1}}$, in the southern group 1.0
  Jy$\mathrm{~beam^{-1}~km~s^{-1}}$, and increase by factors of $\sqrt{2}$. }
\end{figure*}

\begin{figure}[!htb] 
\centering
\includegraphics[width=\columnwidth]{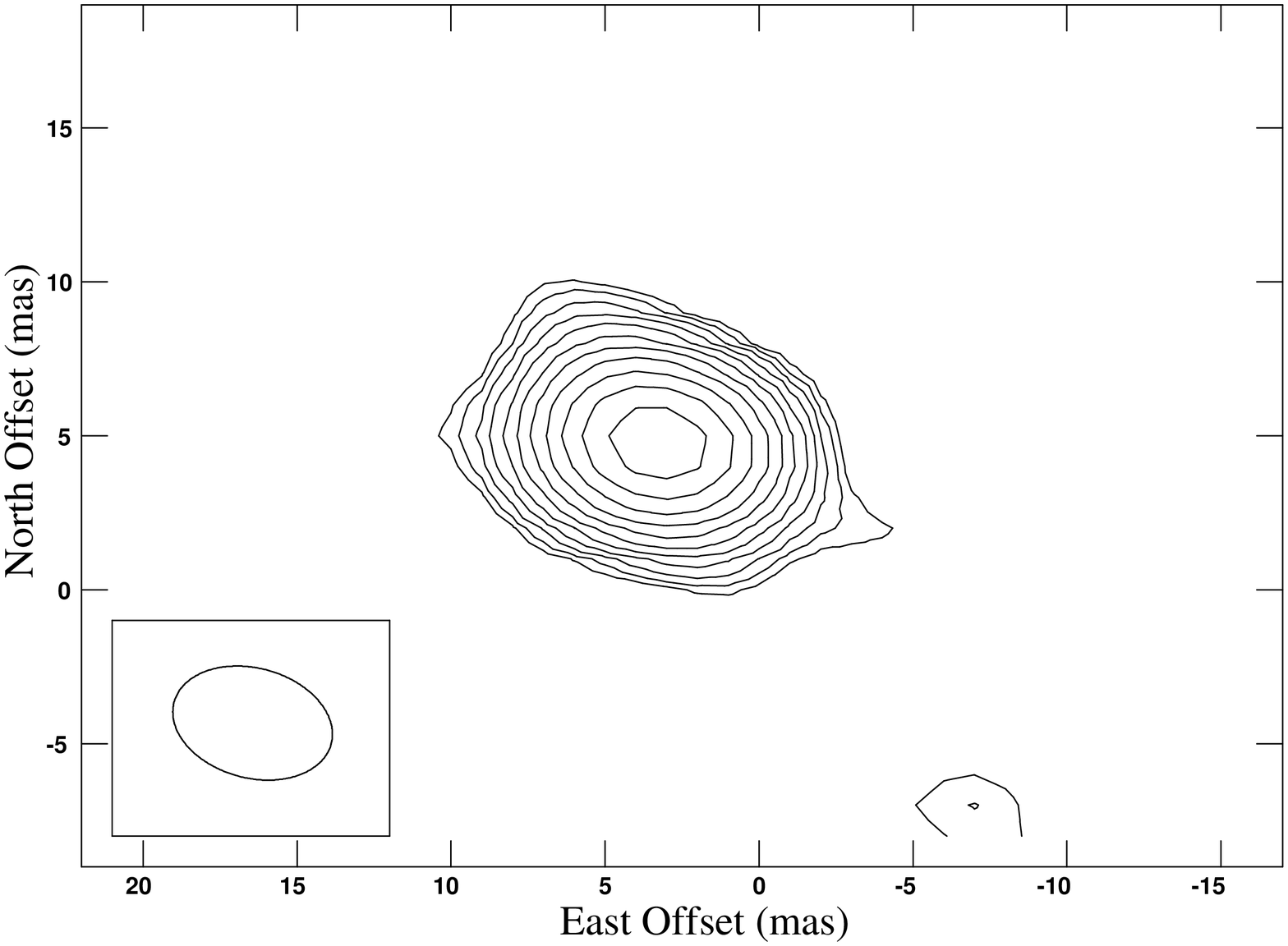}
\includegraphics[width=\columnwidth]{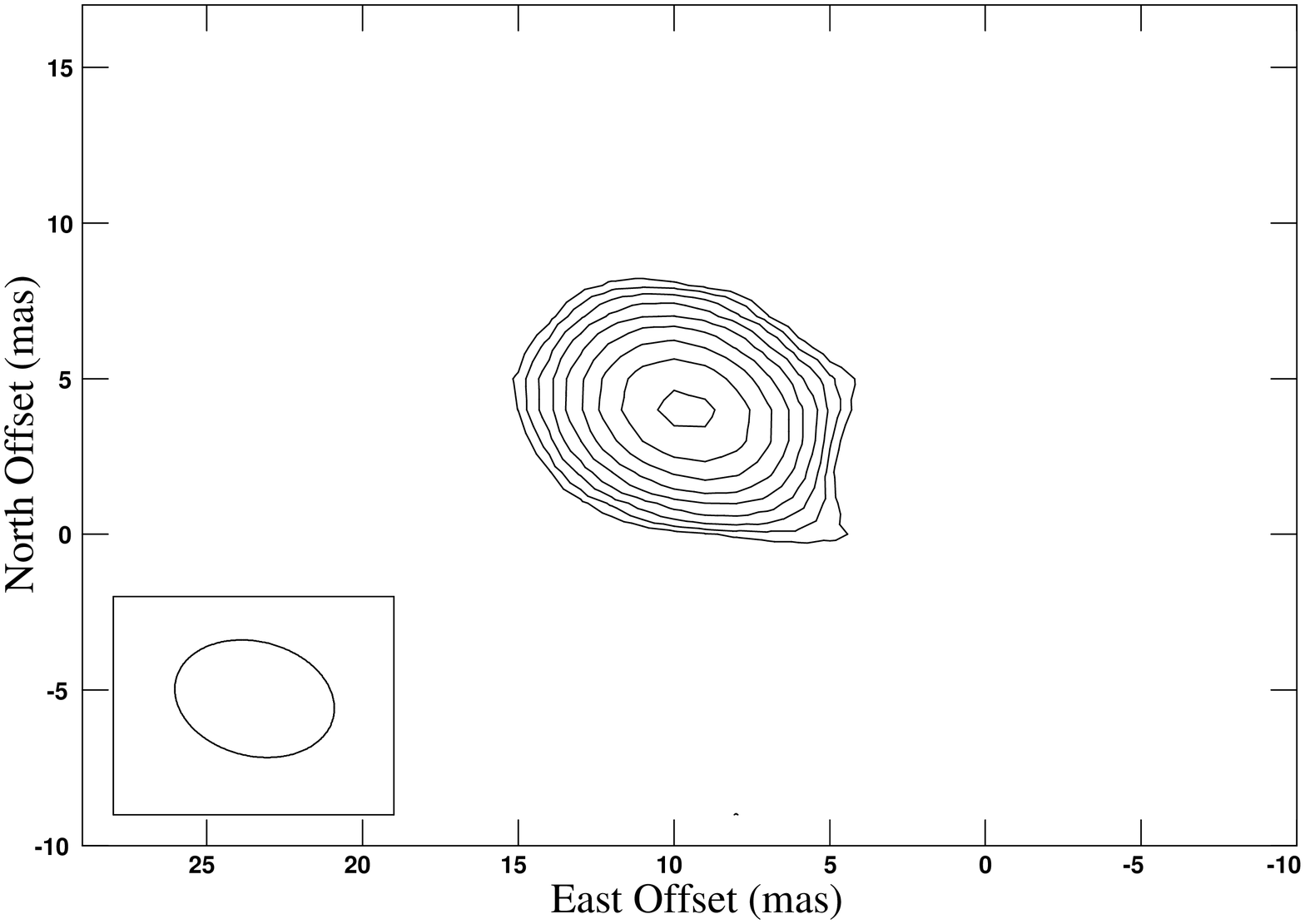}
\caption{Phase-referenced images for two background sources belonging to
  maser ON\,1, J2003+3034 (top) and J2009+3049 (bottom) in the third epoch.  Position offset (0,0) corresponds to the position listed in Table \ref{ta:sour}. The contour levels start at a 3$\sigma$ level, 6.1, and 0.6 mJy beam$^{-1}$, respectively, and increase by factors of $\sqrt2$.}
\end{figure}

\begin{figure*}[!htb] 
\centering
\includegraphics[width=12cm]{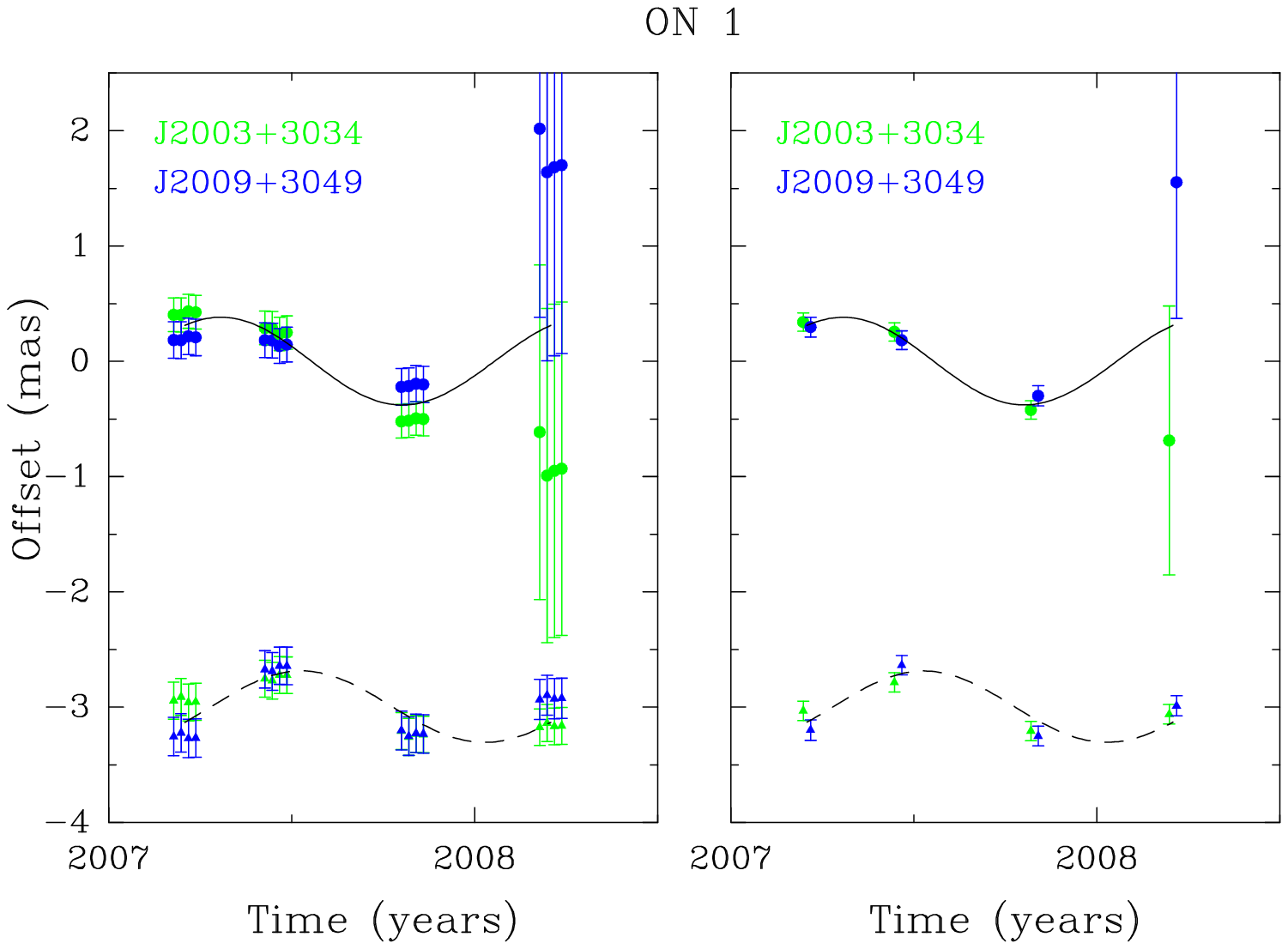}
\caption{\label{fig:on1-pi}Results of the parallax fit based on four maser
  spots. The left graph shows a combined fit on all data, the right graph
  is a fit on the averaged data sets. The filled
colored dots mark the data points in right ascension, while the filled colored triangles
mark the declination. The solid line is the resulting fit in right ascension, the dashed
line in declination. Different colors indicate a different background
source. }
\end{figure*}

\begin{figure}[!htb] 
\centering
\includegraphics[width=6cm]{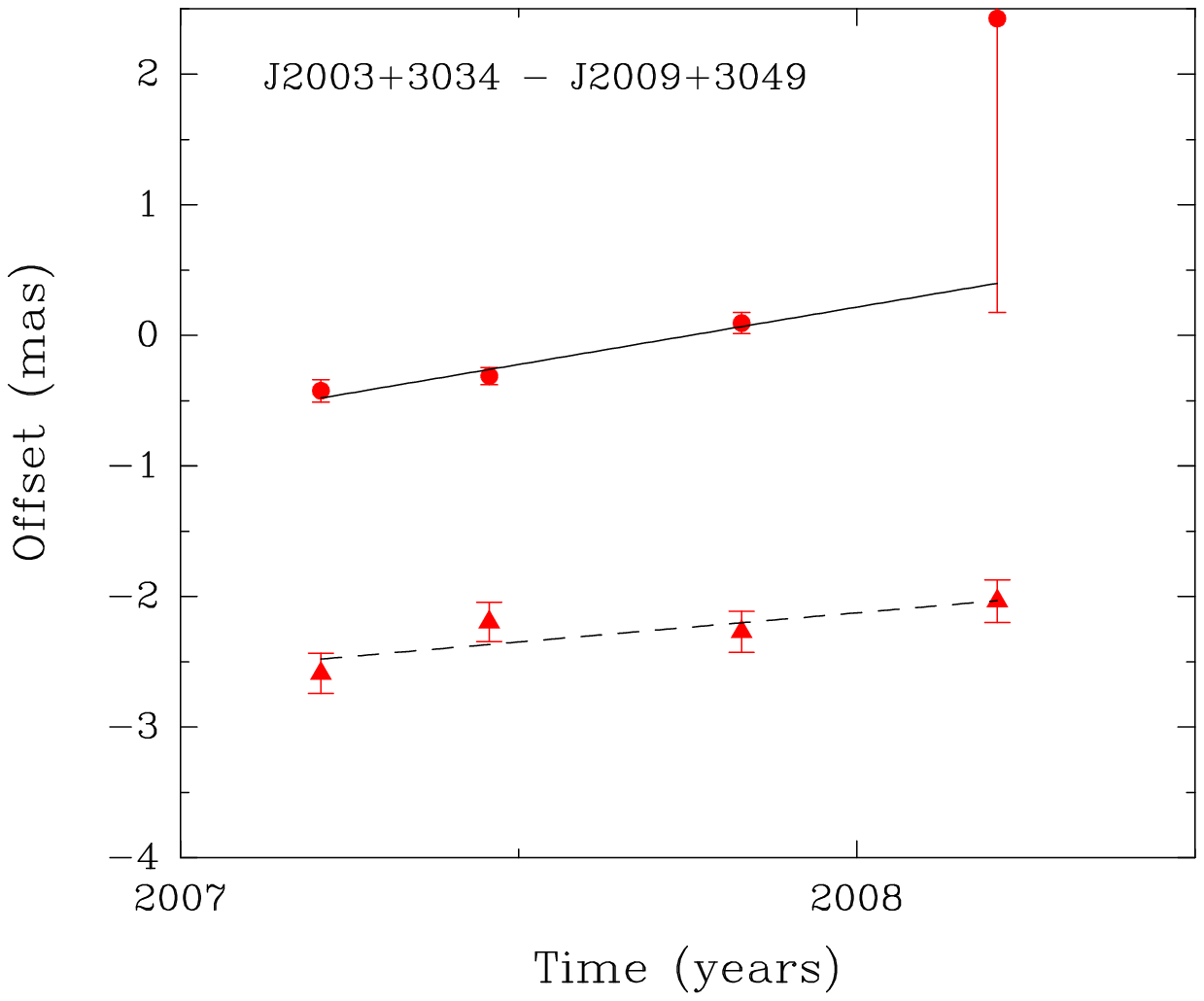}
\caption{\label{fig:on1-q}Variation of the separation between background sources
  J2003+3034--J2009+3049 belonging to ON\,1. The solid line and the dots represent the
  right ascension data, while the dashed line and the filled triangles represent the
  declination data.}
\end{figure}

The maser emission of Onsala 1 (ON\,1) consists of two groups separated
spatially by $\sim$940\,mas, which have different radial velocities by $15\,\mathrm{km~s^{-1}}$ (Fig.~\ref{fig:on1}). The
northern group, which includes the reference channel, is centered at a $v_\mathrm{LSR}$ of
$\approx0\,\mathrm{km~s^{-1}}$ and the southern group at
$\approx15\,\mathrm{km~s^{-1}}$. Four maser spots in the northern group were suitable
for parallax fitting. The masers in the southern group were not used for the parallax fit, since the phase-calibration was less accurate because of the large distance to the phase-reference center (located in the northern maser group). We find a parallax of $0.389\pm0.045$\,mas, corresponding
to a distance of $2.57^{+0.34}_{-0.27}$\,kpc. The proper motions of the southern
maser spots were fitted by assuming the parallax result of the northern group.
All results are listed in Table \ref{ta:data} and the parallax fit is
plotted in Fig.~\ref{fig:on1-pi}. 

For ON\,1, two background sources were detected in epochs two through five. The last
(fifth) epoch had poor $(u,v)$ coverage, which resulted in very elongated
synthesized beams in the east-west
direction (J2003+3034, $12.1\times3.7$\,mas$^2$, P.A.$\sim87\degr$; J2009+3049,
$11.0\times3.3$\,mas$^2$, P.A.$\sim88\degr$) compared to the representative numbers listed in Table \ref{ta:sour}. As a result, the right ascension data had a large uncertainty in epoch five. We estimated this uncertainty from fitting the
variation in the background-source pair separation to be 2\,mas (see Fig.~\ref{fig:on1-q}). Since this was a
combined error for both the background sources, the position error for each
individual background source was $\sqrt{2}$\,mas. We added an additional error floor of $\sqrt{2}$\,mas
to the error given for all the right ascension data points in the fifth epoch.
The apparent movements between the two background sources,
J2003+3034--J2009+3049, were $0.88\pm0.06\,\mathrm{mas~yr^{-1}}$ in right
ascension and $0.45\pm0.21\,\mathrm{mas~yr^{-1}}$ in declination (Fig.~\ref{fig:on1-q}).

\subsection{L\,1206}

\begin{figure}[!htb] 
\centering
\includegraphics[height=0.95\columnwidth,angle=-90]{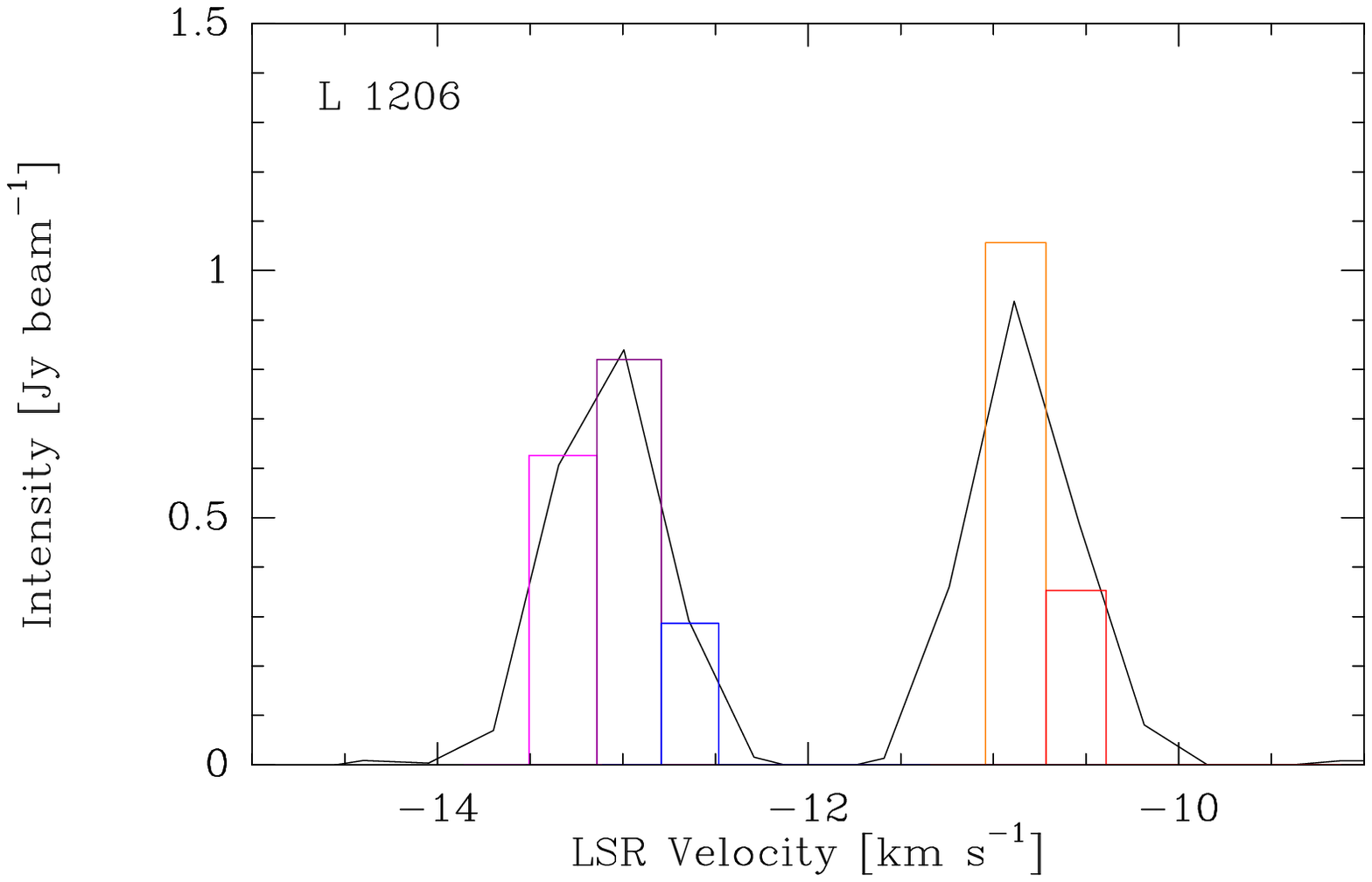}
\includegraphics[width=\columnwidth]{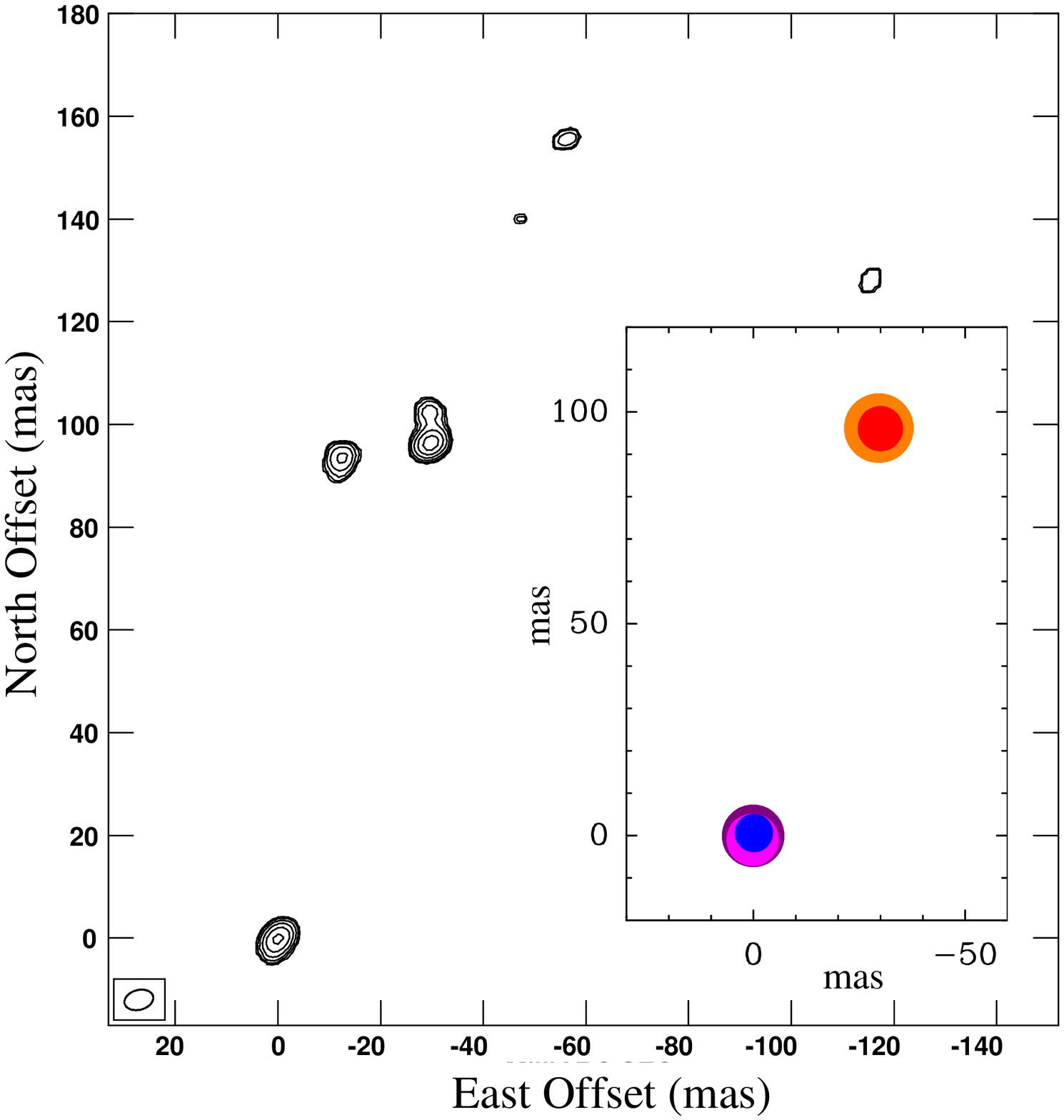}
\caption{\label{fig:l1206} Velocity-integrated map and spectrum of
  L\,1206. Position offset (0,0) corresponds to the position listed in Table \ref{ta:sour}. Maser spots are indicated with color codes for different
    radial velocities. The areas of the colored circles and the colored histogram entries in the spectrum are scaled to the peak flux of that spot. The black line in the spectrum is the intensity of the
    maser within a selected surface, which is not necessarily the same as the
    intensity of the maser spot retrieved from a Gaussian fit. The diffuse
  and weak spots were omitted, such as the spot at (-10,95). Contour levels start at 0.1
  Jy$\mathrm{~beam^{-1}~km~s^{-1}}$ and increase
  by factors of $\sqrt{2}$. }
\end{figure}

\begin{figure}[!htb] 
\centering
\includegraphics[width=\columnwidth]{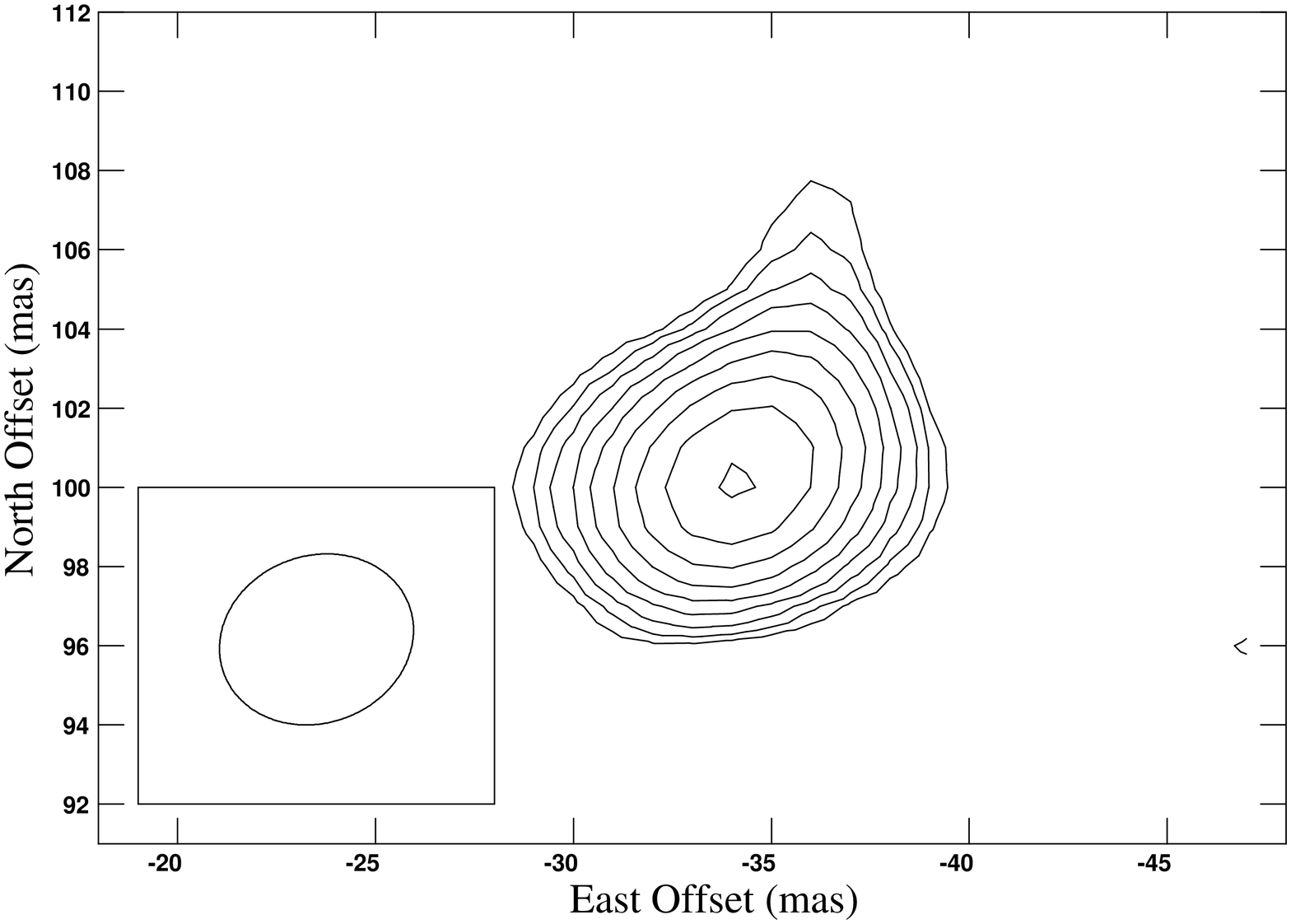}
\includegraphics[width=\columnwidth]{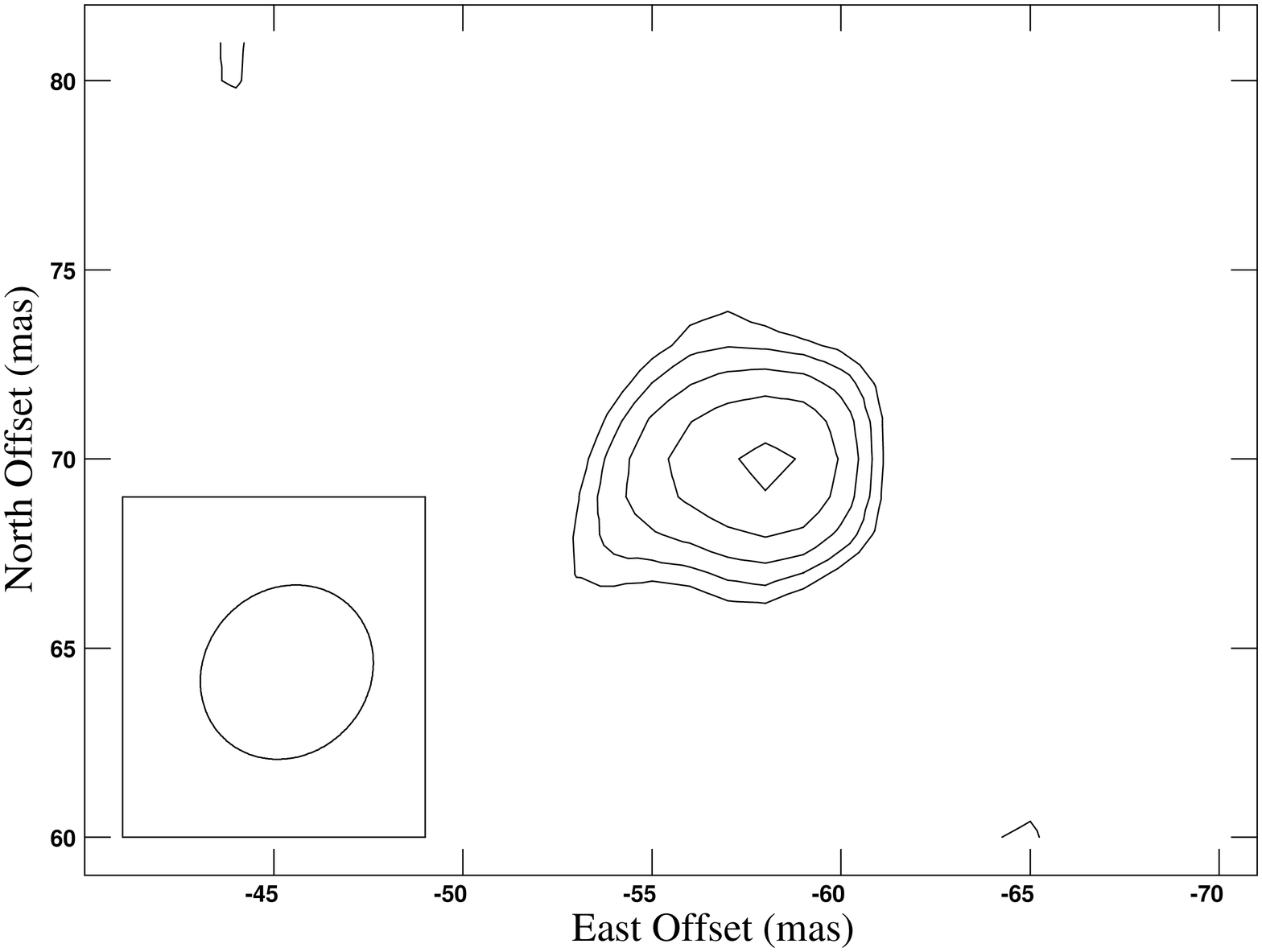}
\caption{Phase-referenced images for two background sources belonging to
  maser L\,1206, J2223+6249 (top) and J2225+6411 (bottom) in the fifth epoch. Position offset (0,0) corresponds to the position listed in Table \ref{ta:sour}. Contour levels start at a 3$\sigma$ level, 3.5, and 0.5
mJy beam$^{-1}$ respectively, and increase by factors of $\sqrt2$.}
\end{figure}

\begin{figure*}[!htb]  
\centering
\includegraphics[width=12cm]{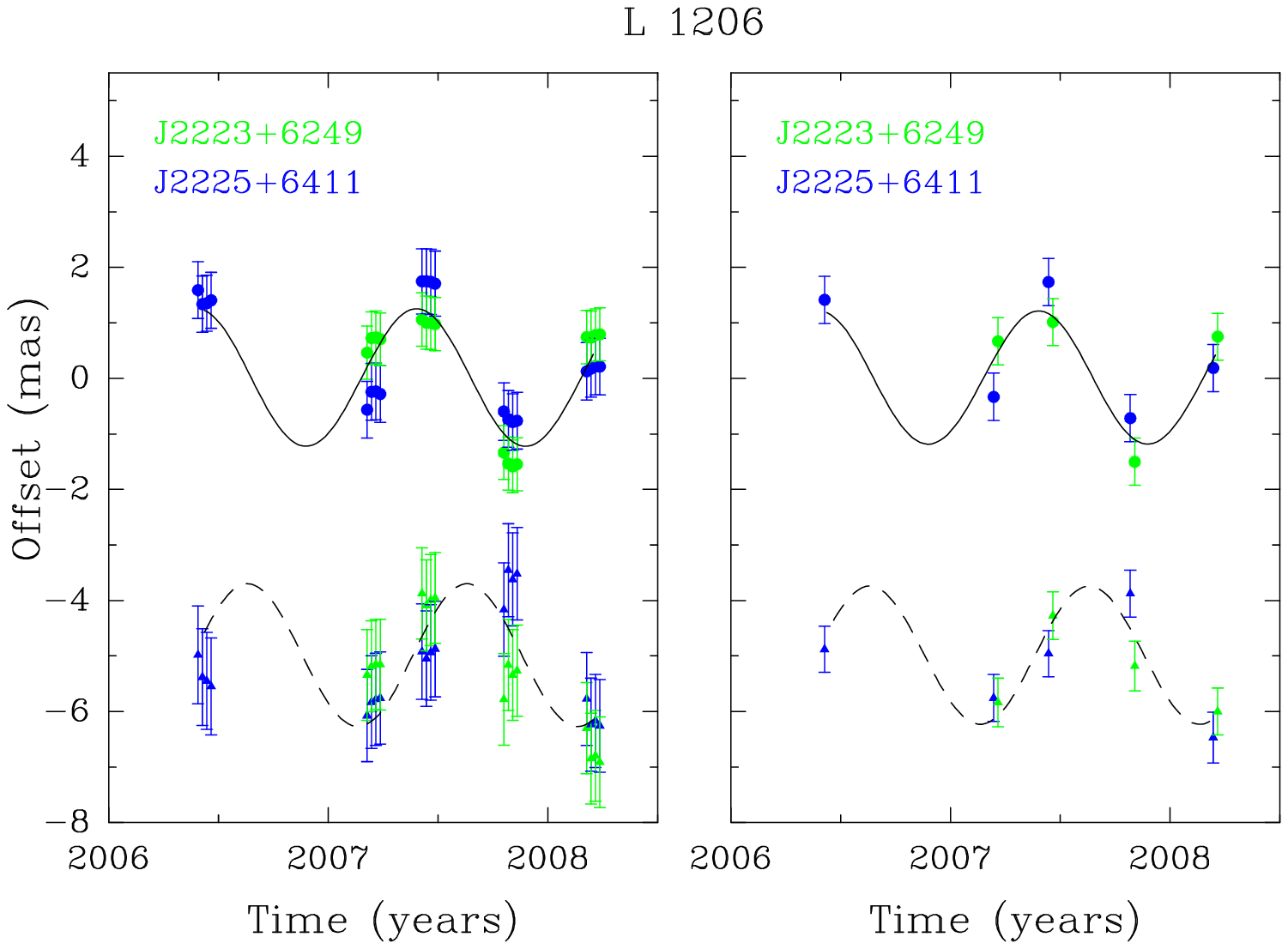}
\caption{\label{fig:l1206-pi}Results of the parallax fit for L\,1206 based on
  four maser spots. The left graph shows a combined fit on all data, the
  right graph is
  a fit on the averaged data sets. The filled
dots mark the data points in right ascension, while the filled triangles mark the
declination. The solid line is the resulting fit in right ascension, the dashed line in
declination. Different colors indicate different background sources.} 
\end{figure*}
\begin{figure}[!htb] 
\centering
\includegraphics[width=6cm]{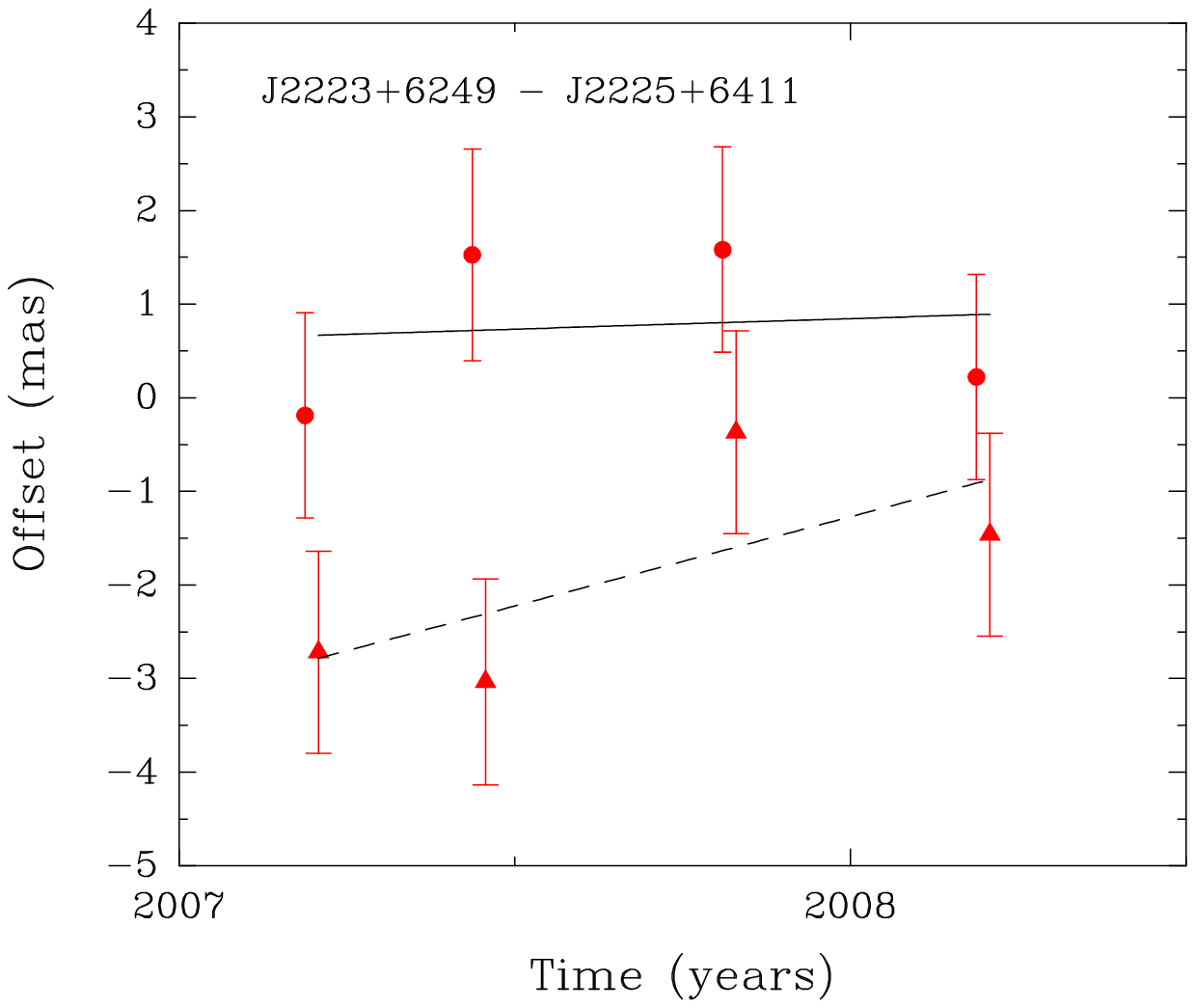}
\caption{\label{fig:l1206-q}Variation of the separation between background sources J2223+6249--J2225+6411
belonging to L\,1206. The solid line and the dots represent the
  right ascension data, while the dashed line and the filled triangles represent the
  declination data.}
\end{figure}

In L\,1206, two maser groups were found separated by $\sim$100\,mas. A
  third maser group was found northeast of the other two. This third maser
  group was not detected in all the epochs because of its weakness so was not used
  in the parallax measurements. Also the maser spot at $-11.6$\,km~s$^{-1}$ was omitted for the same reason. The masers in L\,1206 are shown with a spectrum in Fig.~\ref{fig:l1206}.  The parallax fit used four compact maser spots and resulted in $1.289\pm0.153$\,mas, corresponding
to a distance of $0.776^{+0.104}_{-0.083}$\,kpc. The results of the parallax and proper motion
fits are displayed in Fig.~\ref{fig:l1206-pi} and listed in Table
\ref{ta:data}. 

Only one of the two background sources, J2225+6411, was detected in the
first epoch. The other background source, J2223+6249, had a much larger separation from the maser (i.e., the phase reference)
and the transfer of the phase solutions to the J223+6249 data probably failed. The apparent movements between the background sources were
$0.22\pm1.45\,\mathrm{mas~yr^{-1}}$ in right ascension and
$1.91\pm1.43\,\mathrm{mas~yr^{-1}}$ in declination (Fig.~\ref{fig:l1206-q}).

\subsection{L\,1287}

\begin{figure}[!htb] 
\centering
\includegraphics[height=0.95\columnwidth,angle=-90]{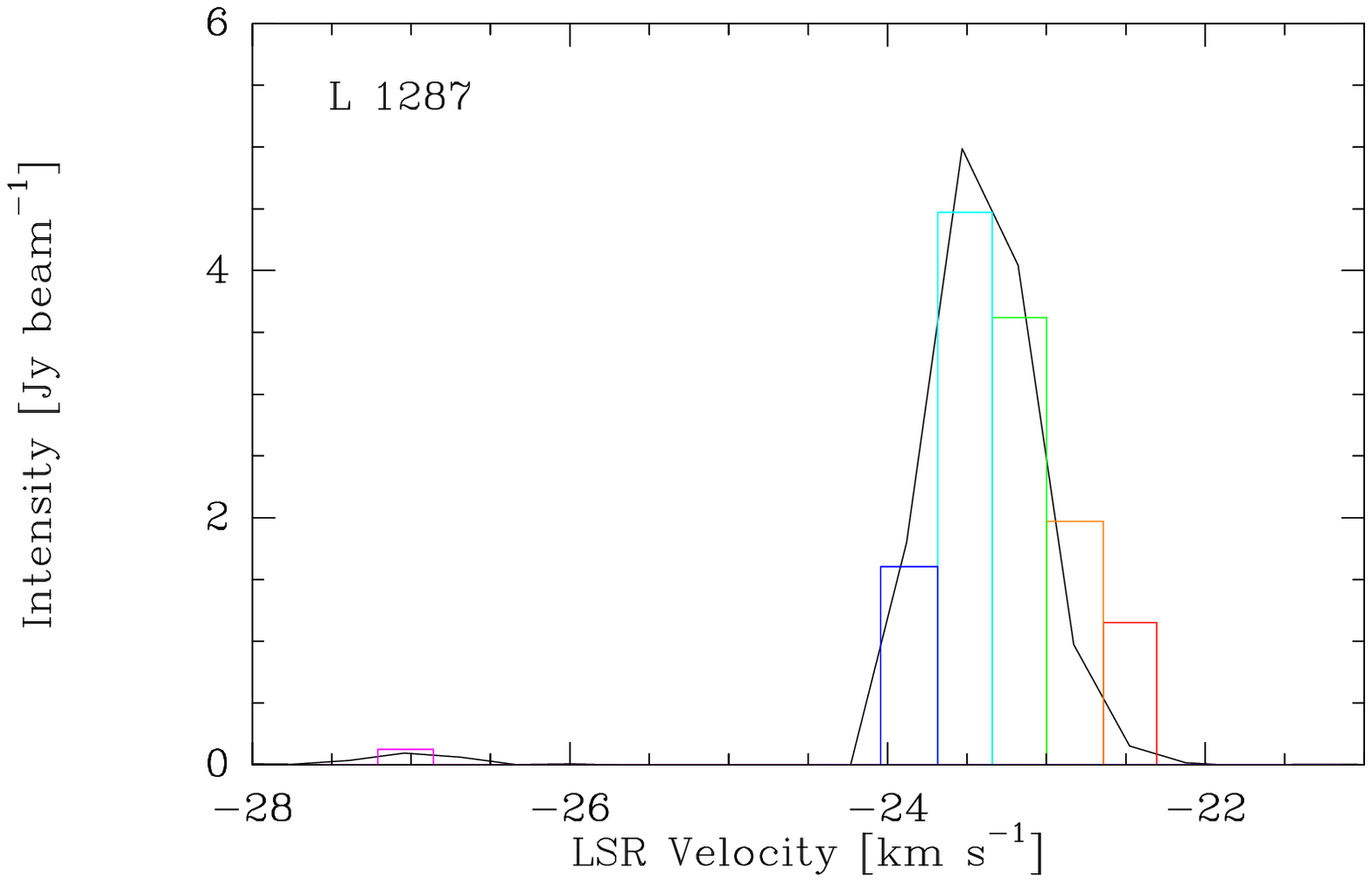}
\includegraphics[width=\columnwidth]{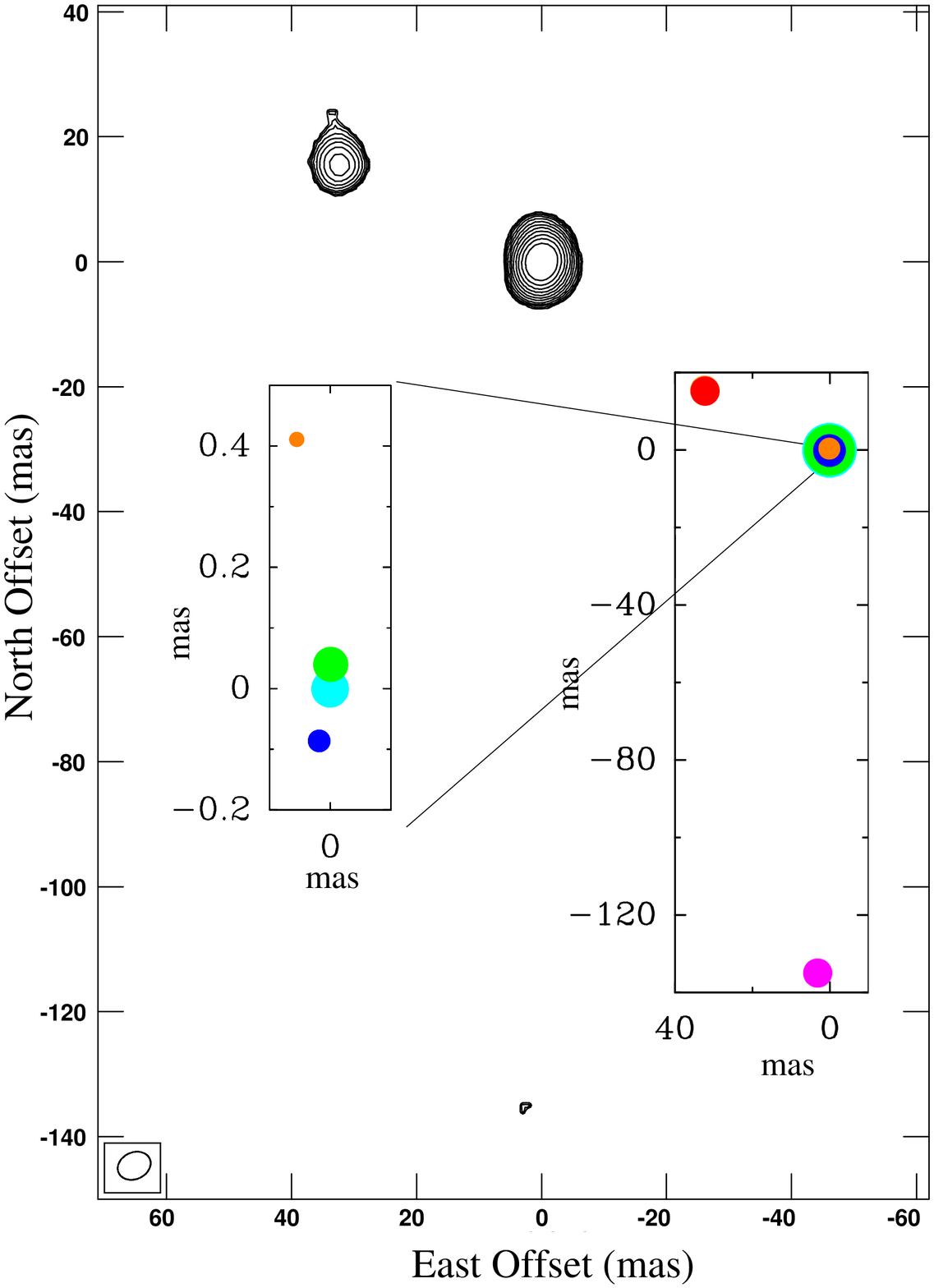}
\caption{Velocity-integrated map and spectrum of
  L\,1287. Position offset (0,0) corresponds to the position listed in Table \ref{ta:sour}. Maser spots are indicated with color codes for different
    radial velocities. The areas of the colored circles and the colored histogram entries in the spectrum are scaled to the peak flux of that
    spot. The black line in the spectrum is the intensity of the
    maser within a selected surface, which is not necessarily the same as the
    intensity of the maser spot retrieved from a Gaussian fit. Contour levels start at 0.05
  Jy$\mathrm{~beam^{-1}~km~s^{-1}}$ and increase
  by factors of $\sqrt{2}$.}
\label{fig:l1287}
\end{figure}

\begin{figure}[!htb] 
\centering
\includegraphics[width=\columnwidth]{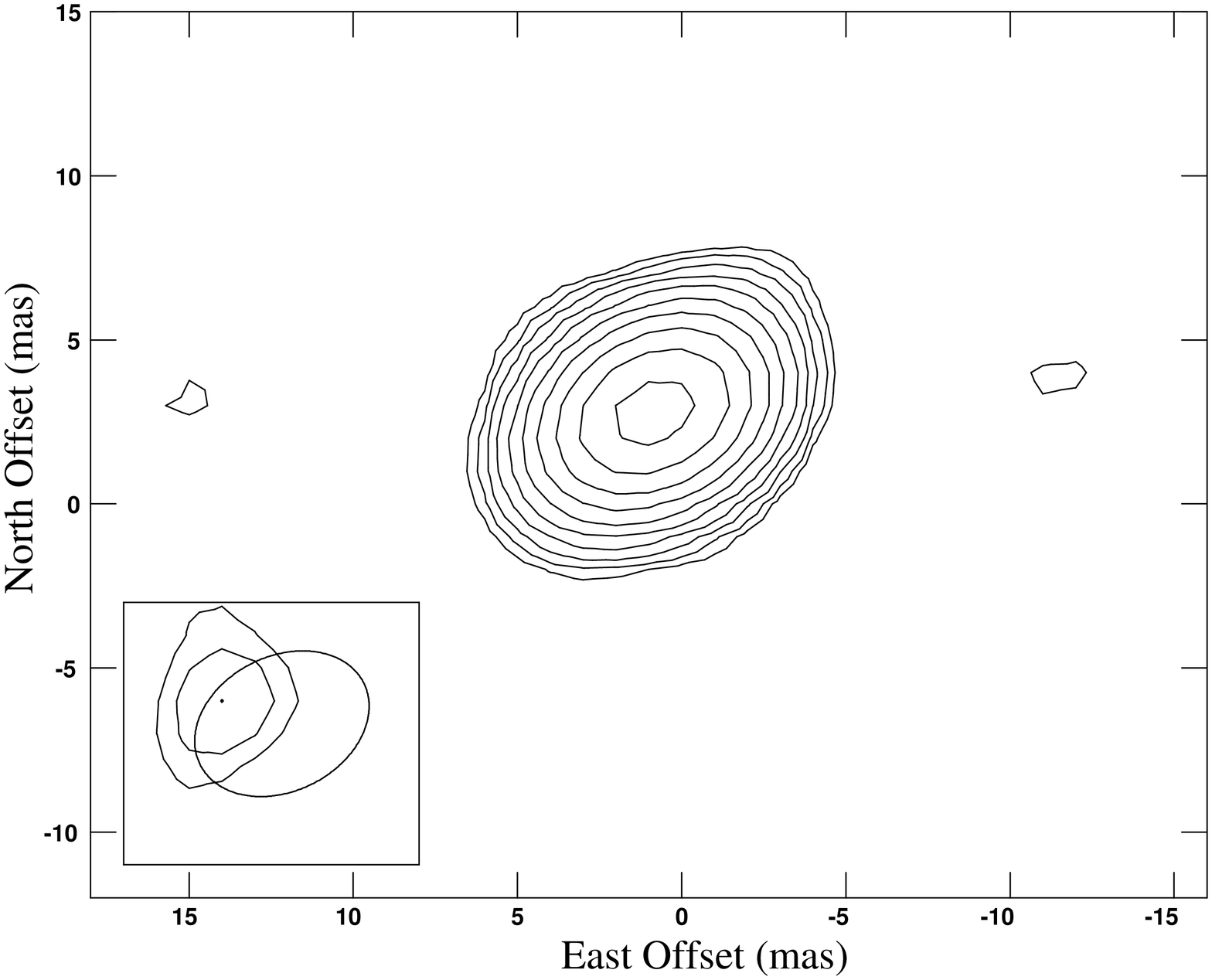}
\includegraphics[width=\columnwidth]{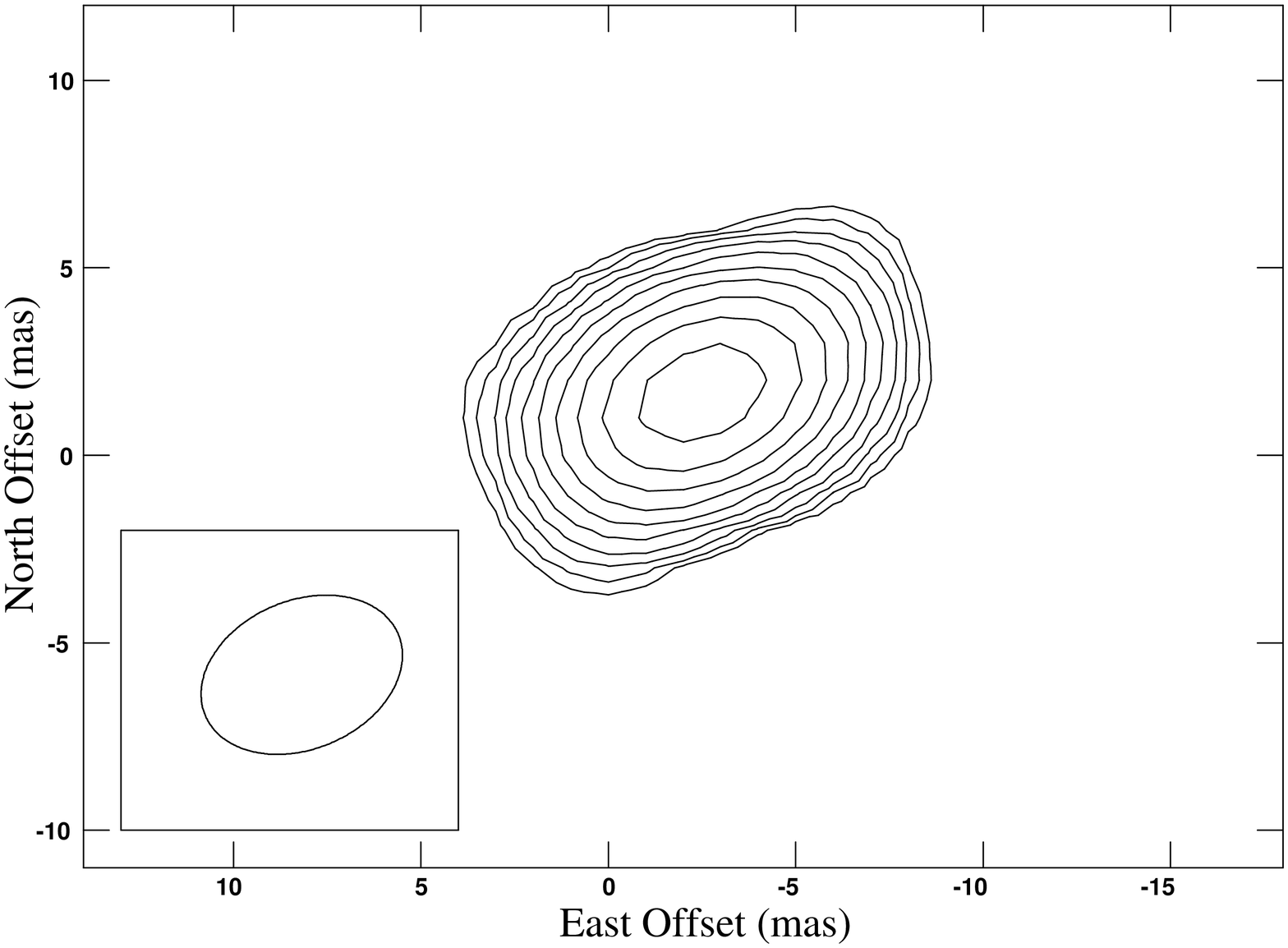}
\caption{Phase-referenced images for two background sources belonging to
  maser L\,1287, J0035+6130 (top)
and J0037+6236 (bottom) in epoch three. Position offset (0,0) corresponds to the position listed in Table \ref{ta:sour}. Contour levels start at a 3$\sigma$ level, 3.6, and 2.4 mJy beam$^{-1}$ respectively, and increase by factors of $\sqrt2$.}
\end{figure}

\begin{figure*} [!htb] 
\centering
\includegraphics[width=12cm]{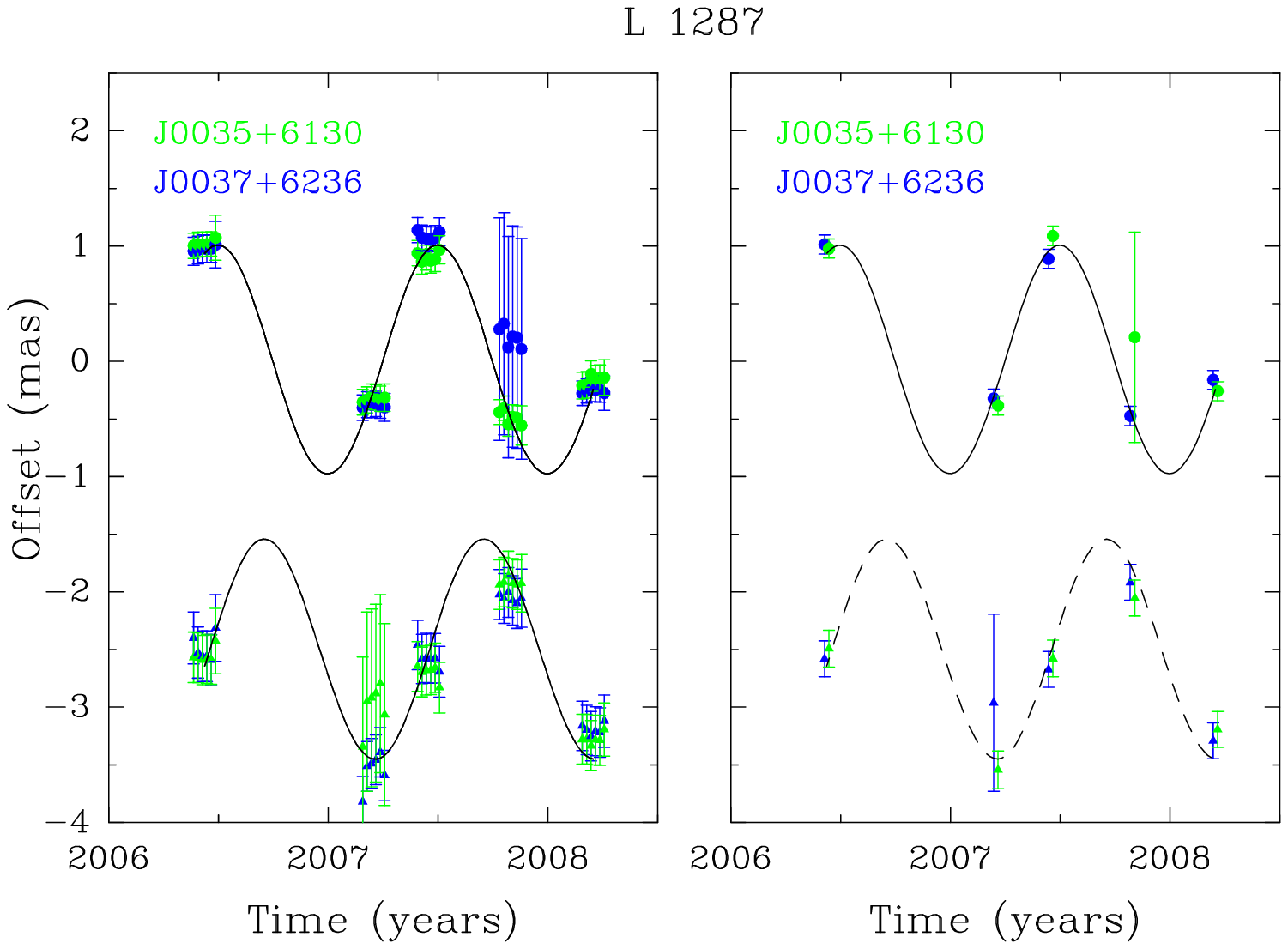}
\caption{Results of the parallax fit for L1278 based on six maser spots. The
  left graph shows a combined fit on all data, while the right graph is a fit on
  the averaged data sets. The filled
dots mark the data points in right ascension, while the filled triangles mark the
declination. The solid line is the resulting fit in right ascension, the dashed line in
declination. Different colors indicate a different background source. }
\label{fig:l1287-pi}
\end{figure*}
\begin{figure}[!htb]  
\centering
\includegraphics[width=6cm]{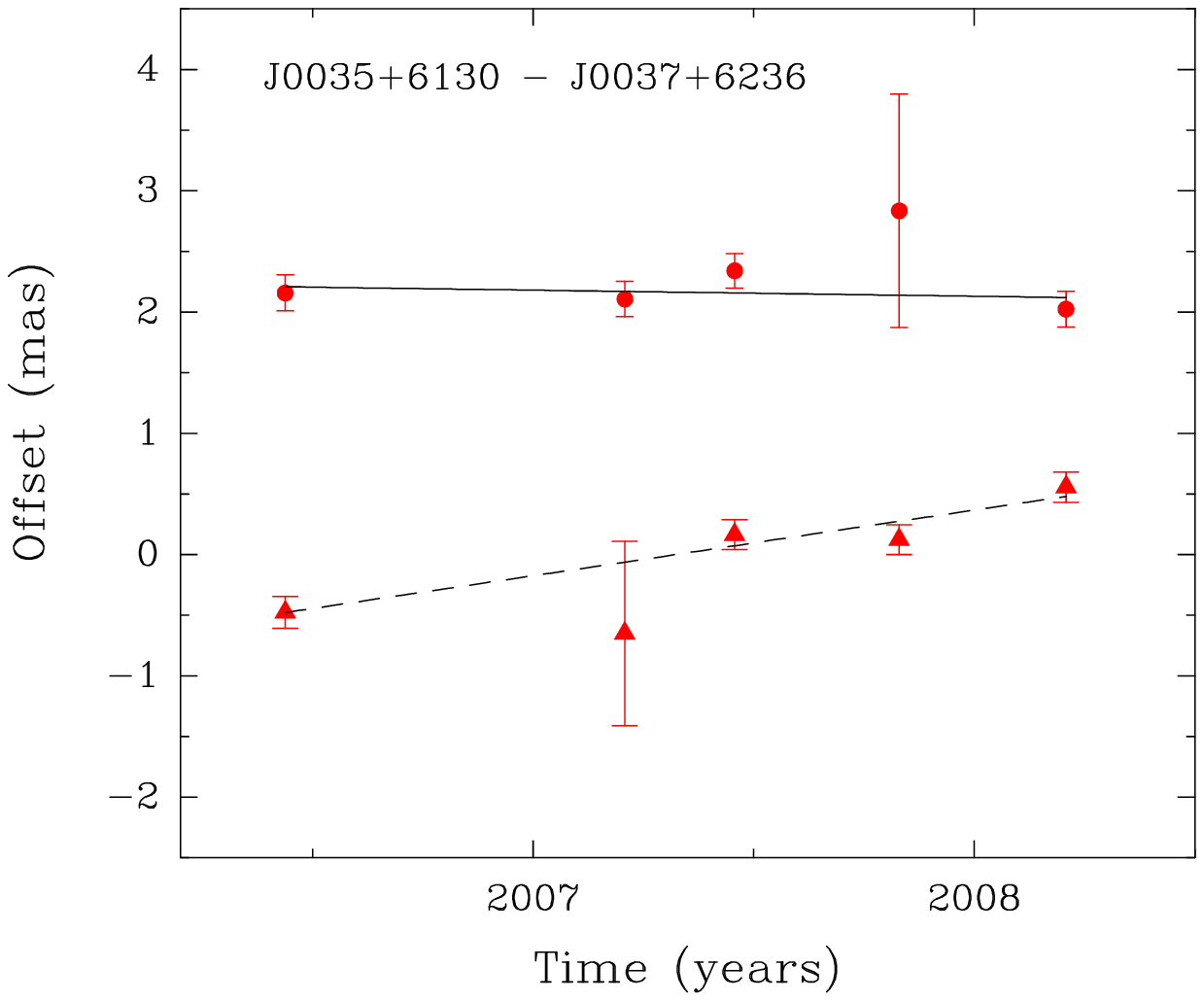}
\caption{Variation of the separation between background sources
  J0035+6130--J0037+6236 belonging to L\,1287. The solid line and the dots represent the
  right ascension data, while the dashed line and the filled triangles represent the
  declination data.}
\label{fig:l1287-q}
\end{figure}

Toward L\,1287 we find three maser groups, two close together (within $\sim40$~mas) and the third $\sim$135
mas southward, shown with a spectrum in Fig.~\ref{fig:l1287}.  A total of six
maser spots from the three groups were used for the parallax fit. We report
a parallax value of $1.077\pm0.039$\,mas, corresponding to a distance of
$0.929^{+0.034}_{-0.033}$\,kpc. The results are plotted in Fig.~\ref{fig:l1287-pi} and
listed in Table \ref{ta:data}. 

The variation of the separation between the two background sources, J0035+6130 and
J0037+6236, is shown in Fig.~\ref{fig:l1287-q}. Two data points appear to be outliers:
the right ascension at epoch four and declination at epoch two.
Increasing the data errors to 0.9 and 0.7\,mas, respectively, for these outliers
yields an apparent movement fit of $-0.05\pm0.12\,\mathrm{mas~yr^{-1}}$ in right
ascension and $0.54\pm0.10\,\mathrm{mas~yr^{-1}}$ in declination (Fig.~\ref{fig:l1287-q}).
We investigated the visibility amplitudes as a function of $(u,v)$ distance and
found evidence of extended structure. For J0037+6236 at the fourth epoch,
evidence of extended structure in the right ascension direction was found; however, for
J0035+6130 at the second epoch, we found no indications of such structure.  

\subsection{NGC\,281-W}

\begin{figure}[!htb] 
\centering
\includegraphics[height=\columnwidth,angle=-90]{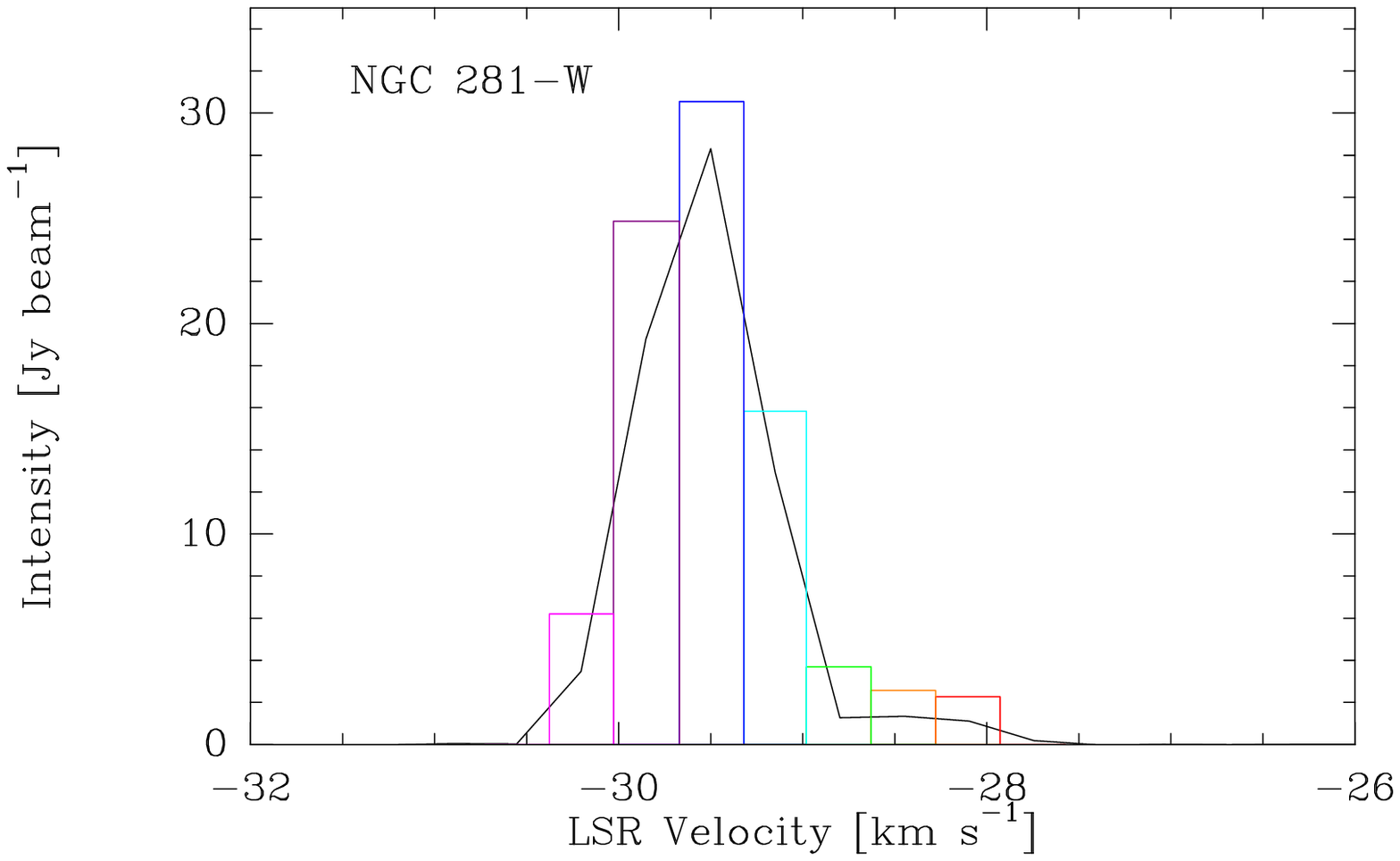}
\includegraphics[width=\columnwidth]{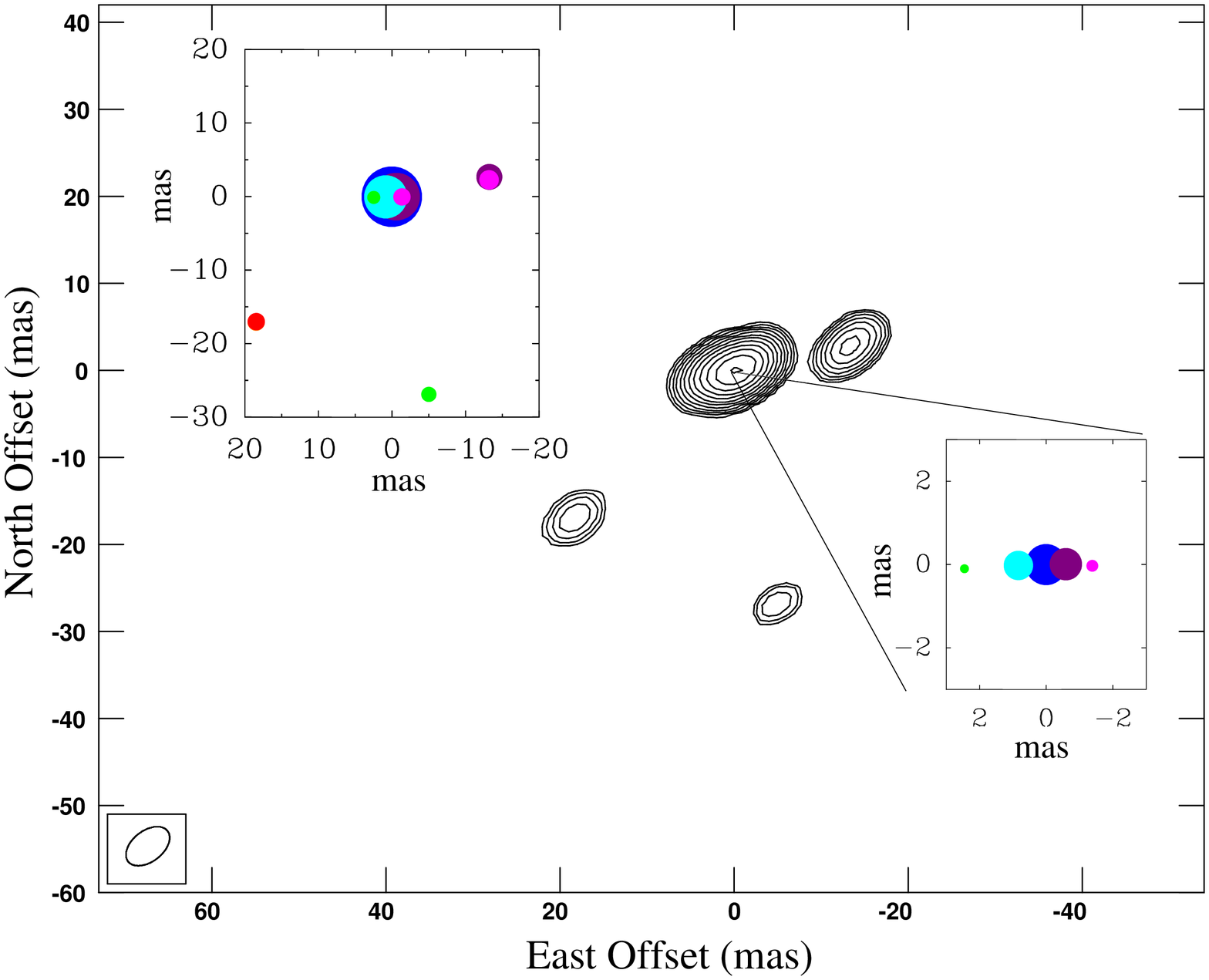}
\caption{\label{fig:ngc}Velocity-integrated map and spectrum of
  NGC\,281-W. Position offset (0,0) corresponds to the position listed in Table \ref{ta:sour}. Maser spots are indicated with color codes for different
    radial velocities. The areas of the colored circles and the colored histogram entries in the spectrum are scaled to the peak flux of that
    spot. The black line in the spectrum is the intensity of the
    maser within a selected surface, which is not necessarily the same as the
    intensity of the maser spot retrieved from a Gaussian fit. Contour levels start at 0.5
  Jy$\mathrm{~beam^{-1}~km~s^{-1}}$ and increase
  by factors of $\sqrt{2}$.}
\end{figure}

\begin{figure}
\centering
\includegraphics[width=\columnwidth]{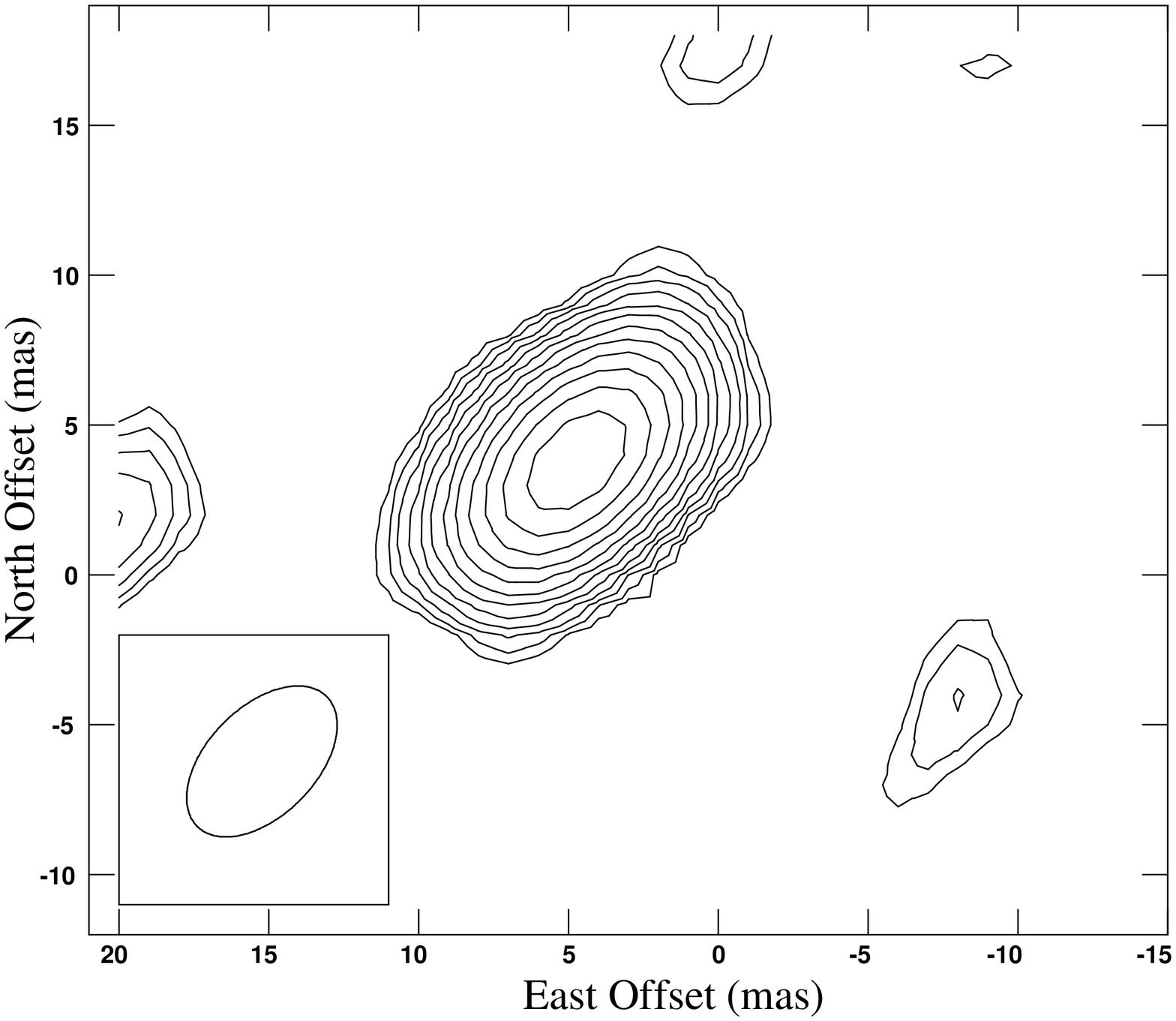}
\includegraphics[width=\columnwidth]{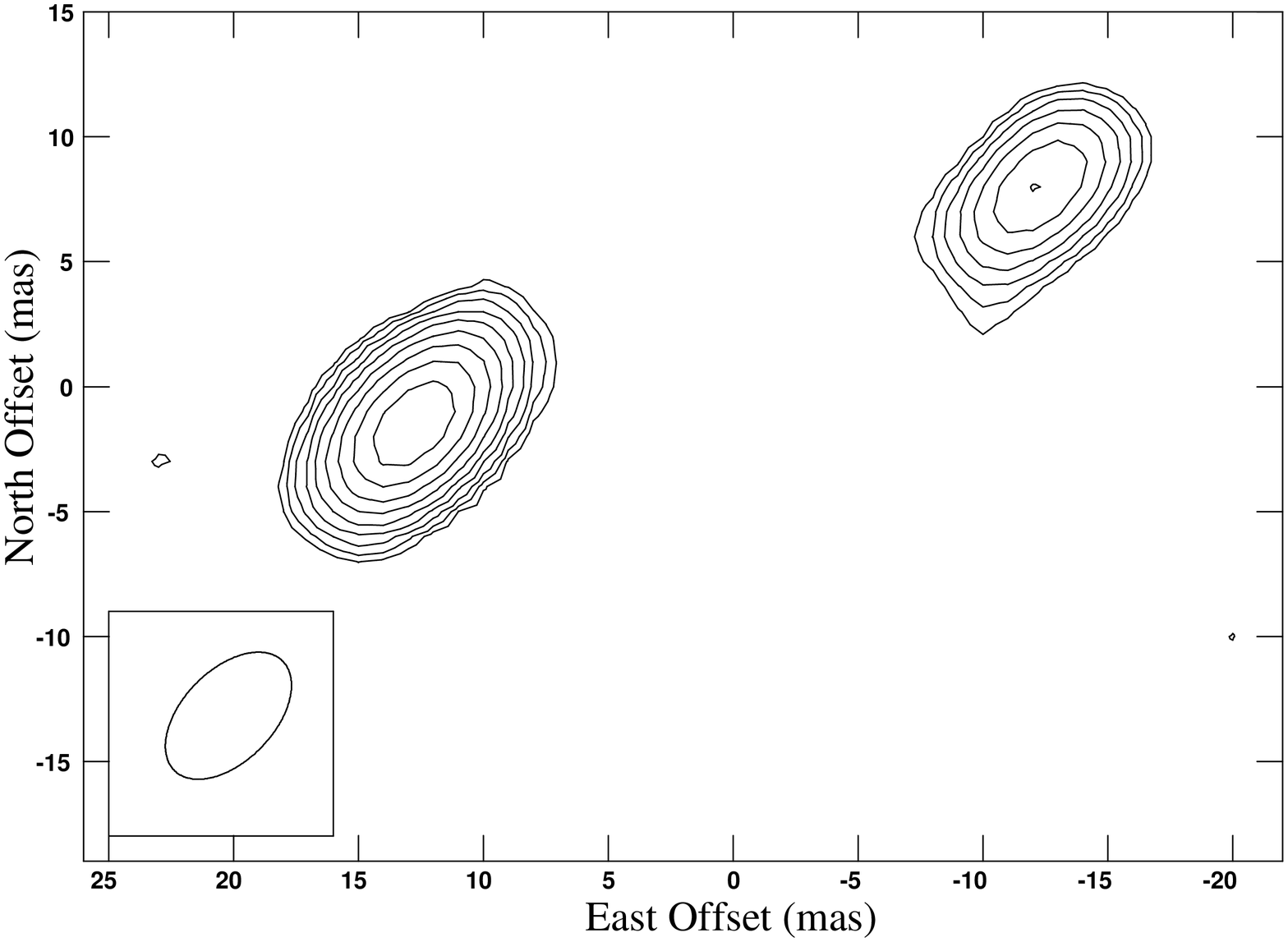}
\caption{Phase-referenced images for two background sources belonging to
  maser NGC\,281-W, J0047+5657 (top) and J0052+5703 (bottom) in epoch five. Position offset (0,0) corresponds to the position listed in Table \ref{ta:sour}. Contour levels start at a 3$\sigma$ level, 2.8, and 0.6
mJy beam$^{-1}$ respectively, and increase by factors of $\sqrt2$. The bottom image shows also the additional component $\sim25$~mas westward of J0052+5703.}
\label{fig:ngc-qmap}
\end{figure}

\begin{figure*}[!htb]  
\centering
\includegraphics[width=12cm]{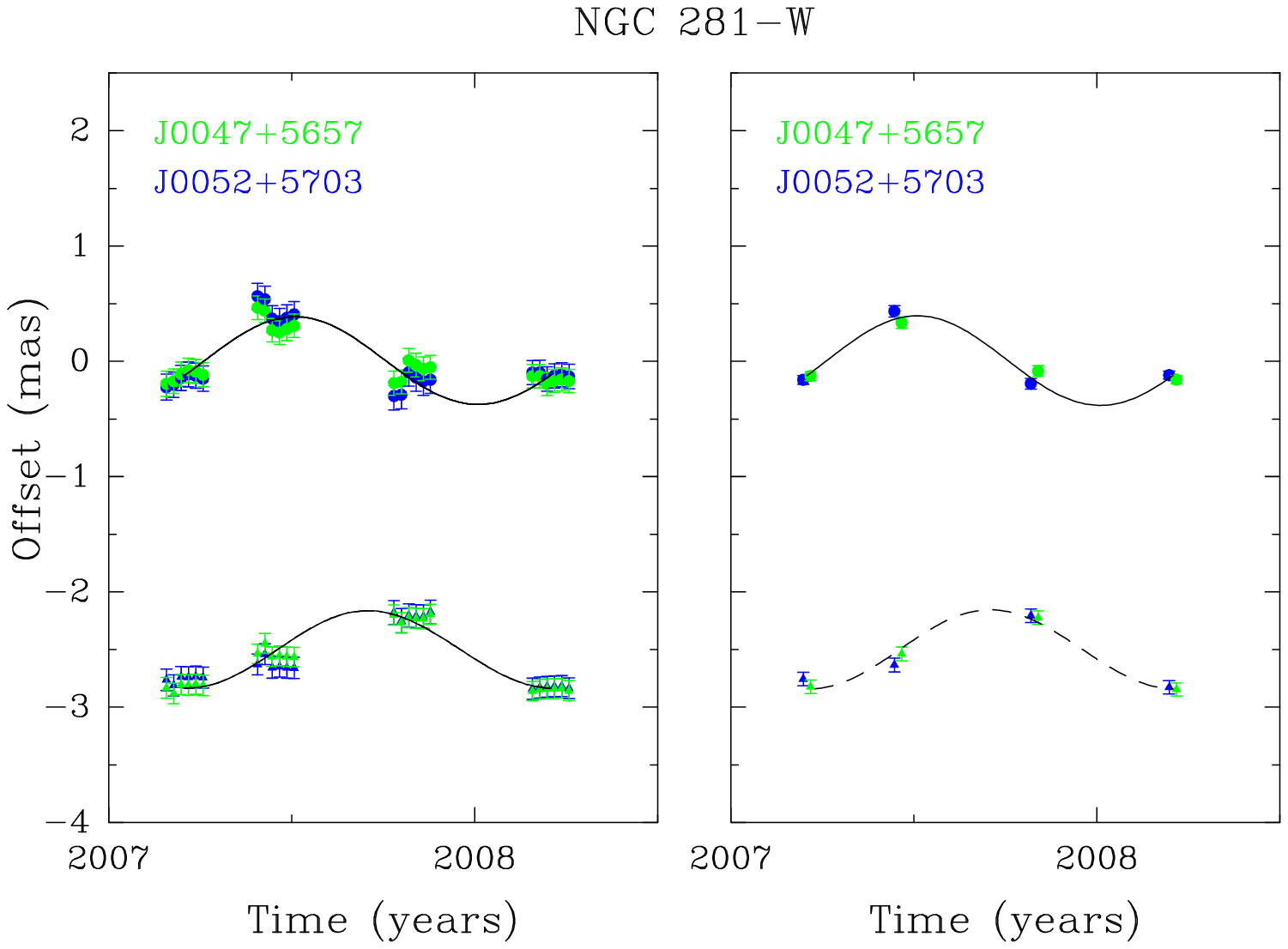}
\caption{\label{fig:ngc-pi} Results of the parallax fit for NGC\,281-W based on
  six maser spots. The left graph shows a fit on all data, while the right
  graph is a fit on the averaged data sets. The filled
dots mark the data points in right ascension, while the filled triangles mark the
declination. The solid line is the resulting fit in right ascension, the dashed line in
declination. Different colors indicate a different background source. }
\end{figure*}
\begin{figure*}[!htb]  
\centering
\includegraphics[width=6cm]{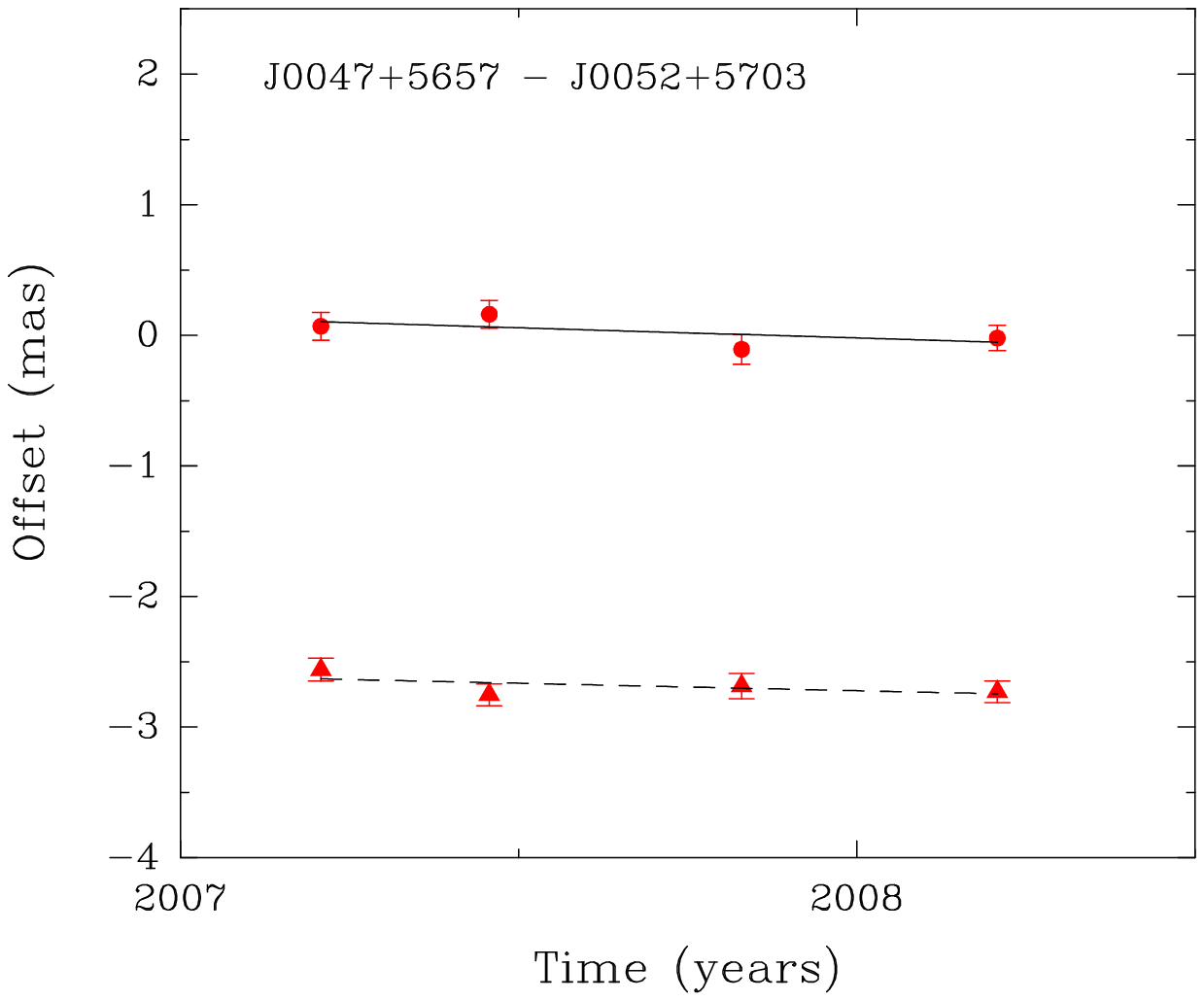}
\includegraphics[width=6cm]{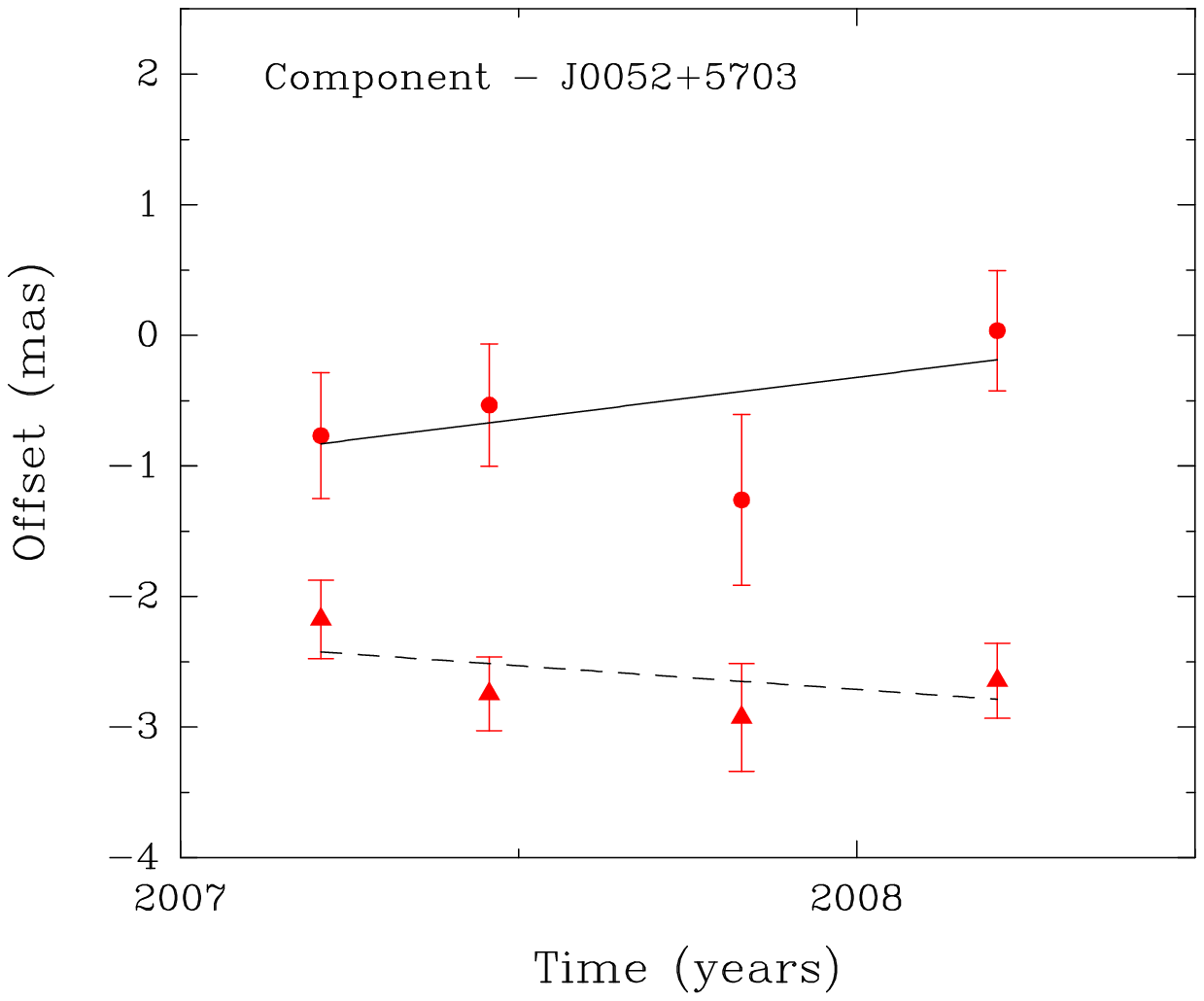}
\caption{{\label{fig:ngc-q}}Left: Variation of the separation between background sources
  J0047+5657--J0052+5703 belonging to NGC\,281-W. Right:
  the proper motion fit on the component of J0052+5703 with respect to the central
  source J0052+5703. The solid line and the dots represent the
  right ascension data, while the dashed line and the filled triangles represent the
  declination data.}
\end{figure*}

We found four groups of methanol masers toward NGC\,281-W, as illustrated in Fig.~\ref{fig:ngc} together with the source spectrum. For the parallax fit we used a total of six maser spots coming from three groups, the central, southern, and southeast groups (Fig.~\ref{fig:ngc}). The weaker maser spots, like the ones located in the western group, were omitted in the parallax fitting. We find a parallax of $0.421\pm0.022$\,mas
corresponding to distance of $2.38^{+0.13}_{-0.12}$\,kpc. The results are shown in
Fig.~\ref{fig:ngc-pi} and listed in Table \ref{ta:data}.

For NGC-281-W, two background sources were detected successfully at epochs
two through five. Background source J0052+5703 showed a double structure with the weaker component
located $\sim$25\,mas westward of the stronger component (Fig.~\ref{fig:ngc-qmap}). The proper motion
of the western component with respect to J0052+5703 was
$\mu_\alpha=0.64\pm0.63\,\mathrm{mas~yr^{-1}}$ and
$\mu_\delta=-0.36\pm0.39\,\mathrm{mas~yr^{-1}}$ (Fig.~\ref{fig:ngc-q}),
i.e, exhibiting no detectable proper motion. The apparent movement between the
two background sources, J0047+5657 and J0052+5703, was also close to zero:
$-0.16\pm0.13\,\mathrm{mas~yr^{-1}}$ in right ascension and
$-0.11\pm0.11\,\mathrm{mas~yr^{-1}}$ in declination.  

\subsection{S\,255}

\begin{figure}[!htb] 
\centering
\includegraphics[width=\columnwidth]{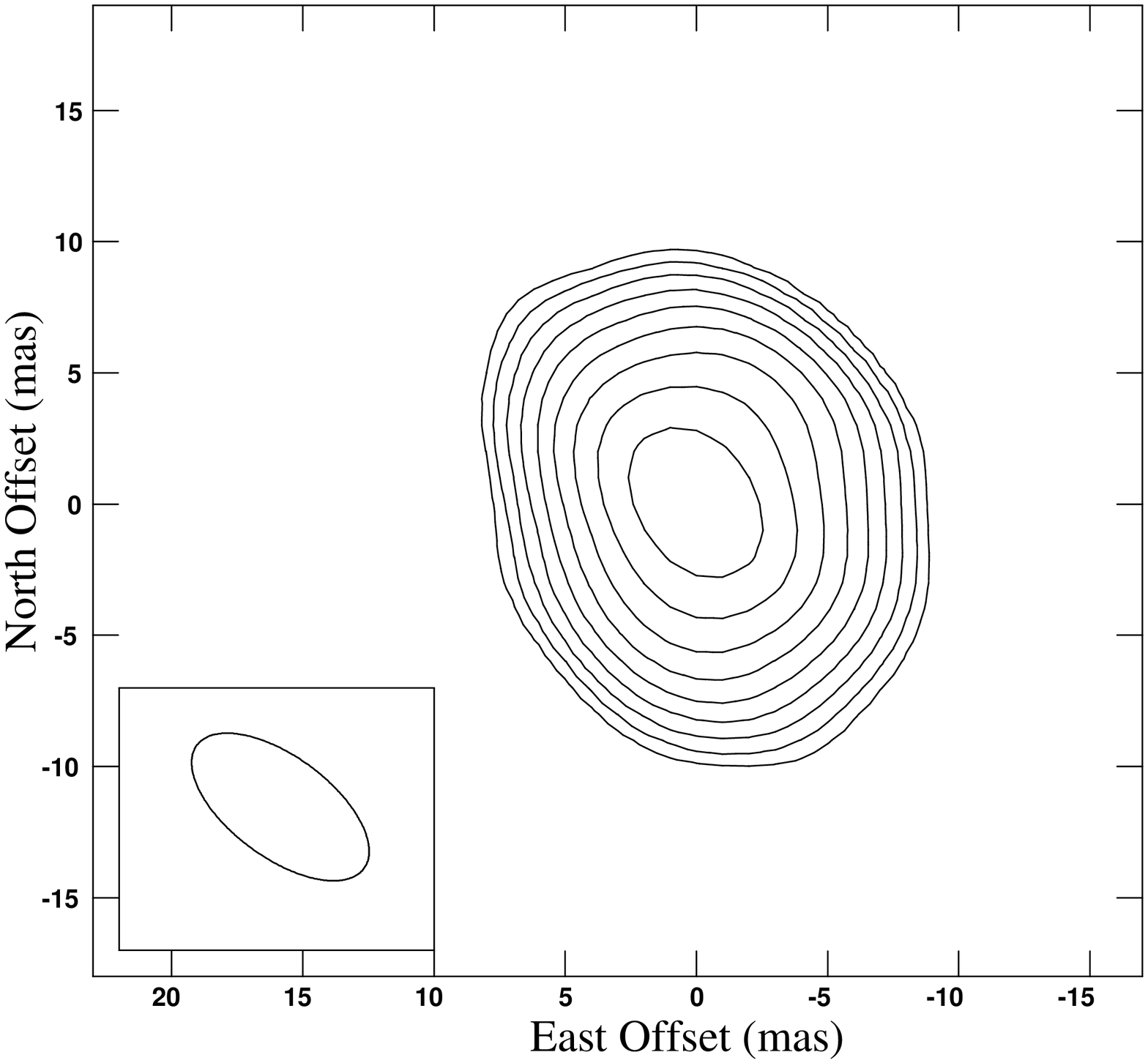}
\caption{\label{fig:s255} Phase-referenced image for maser S\,255 at channel 4.6 km~s~$^{-1}$. Position offset (0,0) corresponds to the position listed in Table \ref{ta:sour}. Contour levels start at 0.23 Jy beam$^{-1}$, and increase by factors of $\sqrt2$.}
\end{figure}

\begin{figure}[!htb] 
\centering
\includegraphics[width=\columnwidth]{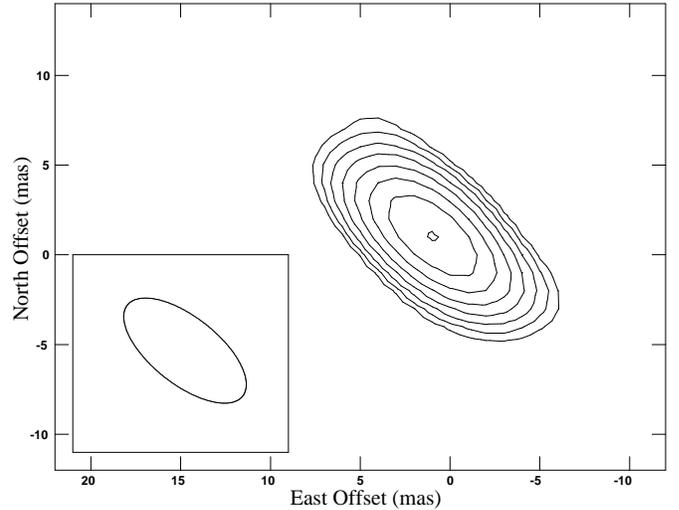}
\caption{\label{fig:s255-q} Phase-referenced images for background source J0613+1708 belonging to
  maser S\,255, in the third epoch. Position offset (0,0) corresponds to the position listed in Table \ref{ta:sour}. The contour levels start at a 3$\sigma$ level, 3.0 mJy beam$^{-1}$, and increase by factors of $\sqrt2$.}
\end{figure}

\begin{figure}[!htb] 
\centering
\includegraphics[width=6cm]{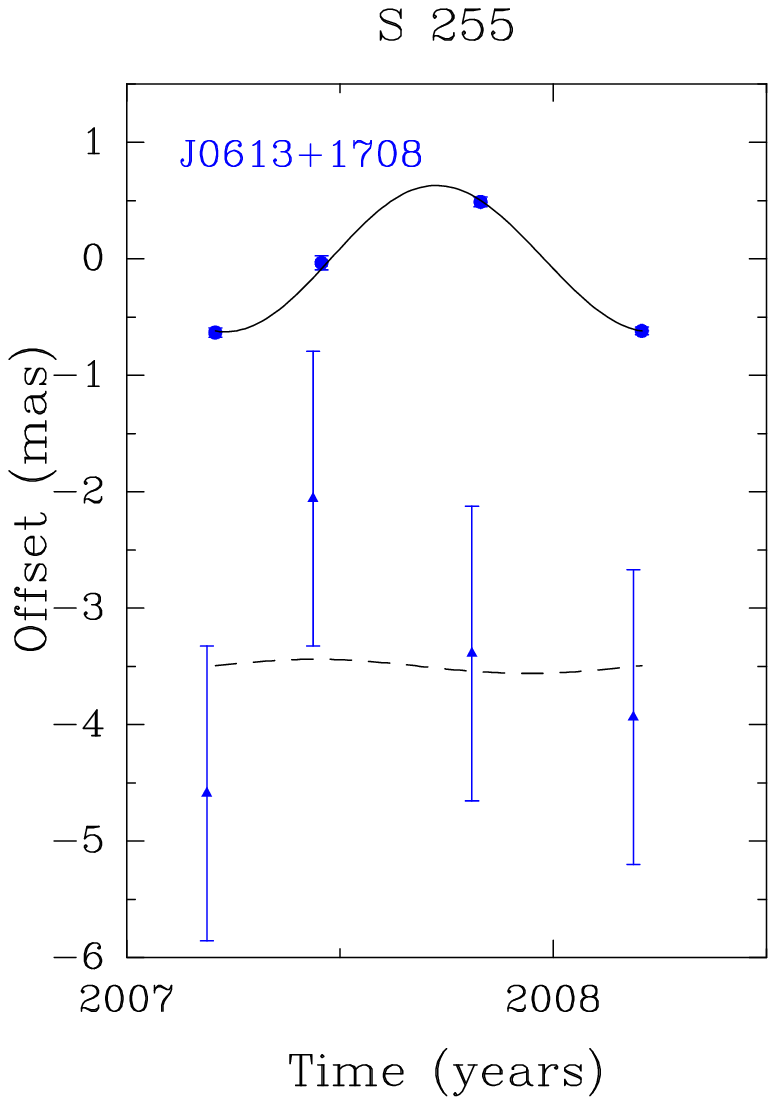}
\caption{\label{fig:s255-pi}Results of the parallax fit for S\,255 based on
  one maser spot at 4.6 km~s~$^{-1}$. The filled
dots mark the data points in right ascension, while the filled triangles mark the
declination. The solid line is the resulting fit in right ascension, the dashed line in
declination.} 
\end{figure}

For S\,255, we could only use emission from one maser channel, namely the reference channel.
Based on this maser spot, at 4.6\,km~s$^{-1}$ (Fig.~\ref{fig:s255}), we find a
parallax of $0.628\pm0.027$\,mas corresponding to a distance of
$1.59^{+0.07}_{-0.06}$\,kpc. The results are shown in Fig.~\ref{fig:s255-pi} and
listed in Table \ref{ta:data}.  

We observed only one background source, J0613+1708, for S\,255 (Fig.~\ref{fig:s255-q}), which was detected in epochs two to five.  With only one
background source, we could not check that its apparent movement was
small, and so, conservatively, we added uncertainties of 1\,$\mathrm{mas~yr^{-1}}$ in both coordinates. 

\section{Discussion}

\subsection{Space motions}

\begin{table*}[!htb]
\caption{\label{ta:3d}Peculiar motions}
\centering
\begin{tabular}{lr@{$\pm$}lr@{$\pm$}lr@{$\pm$}lr@{$\pm$}lr@{$\pm$}lr@{$\pm$}l}
\hline\hline
Source & \multicolumn{6}{c}{$R_0=8.5$ kpc, $\Theta_0=220~\mathrm{km~s^{-1}}$} & \multicolumn{6}{c}{$R_0=8.4$ kpc, $\Theta_0=254~\mathrm{km~s^{-1}}$}\\
       & \multicolumn{2}{c}{$U$} & \multicolumn{2}{c}{$V$} & \multicolumn{2}{c}{$W$} & \multicolumn{2}{c}{$U$} & \multicolumn{2}{c}{$V$} & \multicolumn{2}{c}{$W$}\\
       & \multicolumn{2}{l}{($\mathrm{km~s^{-1}}$)}& \multicolumn{2}{l}{($\mathrm{km~s^{-1}}$)}&\multicolumn{2}{l}{($\mathrm{km~s^{-1}}$)} & \multicolumn{2}{c}{($\mathrm{km~s^{-1}}$)}& \multicolumn{2}{c}{($\mathrm{km~s^{-1}}$)}&\multicolumn{2}{c}{($\mathrm{km~s^{-1}}$)}\\
\noalign{\smallskip}
\hline
\noalign{\smallskip}
%ON\,1    & $13\pm7$ & $-19\pm3$ & $7\pm9$ & &$4\pm7$ & $-20\pm3$& $7\pm9$ \\
ON\,1    & $18$&$8$ & $-19$&3 & $4$& $10$ &$7$&$8$ & $-21$&$3$& $4$&$10$ \\
L\,1206   & $2$&$4$ & $-16$&$2$ & $0$&$6$  &$-1$&$4$ & $-16$&$2$& $0$&$6$\\
L\,1287   & $13$&$2$ & $-18$&$2$ & $-3$&$2$  &$10$&$2$ & $-18$&$2$ &$-3$&$3$\\
NGC\,281-W& $13$&$2$ & $-3$&$2$ & $-9$&$2$  &$6$&$2$& $-3$&$2$ &$-9$&$2$\\ 
S\,255    & $1$&$3$ & $-4$&$12$ & $3$&$7$  & $2$&$3$ & $-4$&$12$ & $3$&$7$\\
\noalign{\smallskip}
\hline
\end{tabular}
\end{table*}

\begin{figure}
\centering
\includegraphics[width=\columnwidth]{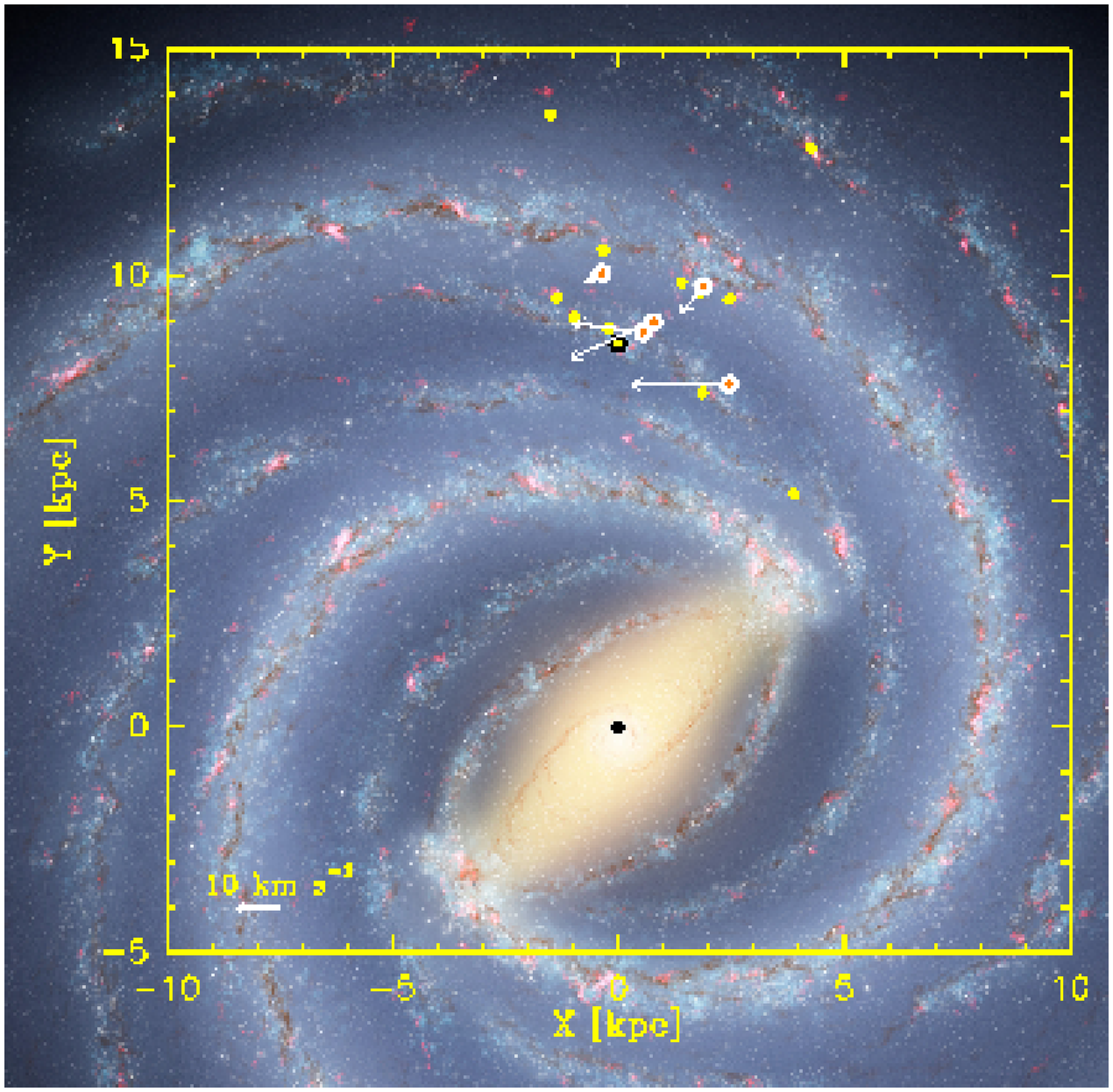}
\caption{\label{fig:gal}An artist's impression of a plane-on view of our Galaxy (image
credit:
 R. Hurt NASA/JPL-Caltech/SSC), overlaid with parallax measurements of
water and methanol
 masers between $60\degr<l<240\degr$ taken from \citet{reid:2009b}, in which recent parallax measurements have been put together to study the Galactic structure. The black dot marks the Galactic center, the yellow dot encircled with black the Sun, the yellow dots mark the parallaxes taken from the literature, the orange dots encircled with white, and white arrows mark the parallaxes and peculiar motions obtained in this work. The peculiar motions are shown after
  removing the Galactic rotation assuming the values of $R_0=8.4$ kpc and $\Theta_0=254~\mathrm{km~s^{-1}}$.}
\end{figure}

With parallax and proper motion measurements, one can calculate the full
space motion of the masers in an SFR in the Galaxy.  Using an accurate model of Galactic
dynamics (Galactic rotation speed, ${\Theta_0}$, distance of the Sun to the Galactic center, ${R_0}$,
rotation curve) for a source and removing the modeled contribution of Galactic rotation allows one to retrieve source peculiar motion relative to a circular orbit.
The peculiar motion is described by the vectors $U$, $V$, and $W$, locally
toward the Galactic center, toward the direction of rotation and toward the
North Galactic Pole, respectively.

In Table \ref{ta:3d}, the peculiar motions of our sources are given for two
different Galactic models: the I.A.U.~recommended values for the LSR
motion $R_0=8.5$ kpc, $\Theta_0=220~\mathrm{km~s^{-1}}$ in columns 2--4; a new model with $R_0=8.4$ kpc, $\Theta_0=254~\mathrm{km~s^{-1}}$, based
on a large sample of trigonometric parallax measurements \citep{reid:2009b},
in columns 5--7. Both models adopt the Hipparcos values for the Solar motion
\citep{dehnen:1998} and assume a flat rotation curve. We used the parallaxes and proper motions as are listed in Table \ref{ta:sum}, where the uncertainty introduced by the apparent movement of the background sources was included in the proper motion uncertainty of the source.

In Fig.~\ref{fig:gal} we plot the SFRs studied in this work at their determined distances with their peculiar motion in the Galactic plane, after removing the Galactic rotation. 
For either model, it is clear that three SFRs show a similar lag on circular
rotation as the SFRs studied by \citet{reid:2009b}, while two have a circular
rotation close to zero lag.  
For NGC\,281-W, the zero lag can be understood from its large distance from the
Galactic disk ($b\simeq-6\degr$) and the dominant contribution from the expanding super bubble to the peculiar motion of this SFR (see Sect.~\ref{sec:ngc}). For S\,255, it is not clear why its
circular velocity is close to Galactic rotation (Sect.~\ref{sec:s255}). 
The motion towards the Galactic center, $U$, tends to be positive for most of the SFRs. 

Recently, \citet{mcmillan:2009} have published a reanalysis of the maser astrometry
presented by \citet{reid:2009b}. They fitted the data with the revised solar
peculiar motion of Binney~(2009), mentioned in \citet{mcmillan:2009}, where the $V$ component of the solar peculiar
velocity, $V_\odot$,
 is increased from 5.2~km~s$^{-1}$ to 11~km~s$^{-1}$.
 With a larger $V_\odot$ the lag of SFRs on Galactic rotation would decrease to $\sim11~\mathrm{km~s^{-1}}$,
which agrees with the expected velocity dispersion for young stars of
$\sim10~\mathrm{km~s^{-1}}$ \citep{aumer:2009}. However, even when considering
the revised $V_\odot$, we note that most SFRs still rotate more slowly than
the Galactic rotation (by $\sim11~\mathrm{km~s^{-1}}$) and that their velocities are not randomly dispersed.

\subsection{Onsala 1}
Our trigonometric parallax measurement places ON\,1 on a distance from the Sun of $2.57^{+0.34}_{-0.27}$\,kpc, somewhat closer than the (near) kinematic distance of 3.0\,kpc, based on the
methanol maser line at $15\,\mathrm{km~s^{-1}}$. The latter assumes a flat rotation curve with as distance of the Sun to the Galactic center, $R_0=8.5$\,kpc, and for the Galactic rotation speed, $\Theta_0=220$\,km~s$^{-1}$. However, it is farther than
the commonly adopted (near) kinematic distance of 1.8\,kpc based on
formaldehyde at $v_\mathrm{LSR}=11.2\,\mathrm{km~s^{-1}}$ \citep{macleod:1998} using the Galactic
rotation curve of \citet{wouterloot:1989}, with  $R_0=8.5$\,kpc,
$\Theta_0=220$~km~s$^{-1}$. Our result resolves the near/far kinematic distance
ambiguity and locates ON\,1 in the Local spur consistent with the near kinematic distance.

We found different proper motions for the northern and southern maser
groups (Table \ref{ta:data}). 
The resulting difference in the east-west direction was small, $\mu_\alpha=0.18\pm0.24\,\mathrm{mas~yr^{-1}}$,
but in the north-south direction, the southern group was moving by $\mu_\delta=-0.77\pm0.12\,\mathrm{mas~yr^{-1}}$ away from the
northern group. At a distance of 2.6\,kpc, this corresponds to a relative speed of 9.4\,km~s$^{-1}$.

This large proper motion may be explained by the masers being located in the
molecular gas surrounding an expanding H{\sc ii} region,
as suggested by \citet{fish:2007} and \citet{su:2009}. The H{\sc ii} region is located at 
($\alpha=20^{\mathrm{h}}10^{\mathrm{m}}09\rlap{$.$}\,^{\mathrm{s}}03,~\delta= +31\degr 31\arcmin 35\rlap{$.$}\,\arcsec 4$, J2000) and between the two maser groups. The radial velocity of the H{\sc ii}
region, from the H76$\alpha$ recombination line, is $5.1 \pm 2.5$~km~s$^{-1}$ \citep{zheng:1985}. 
The methanol masers would be (with respect to the rest frame of ON\,1 at $5.1$\,km~s$^{-1}$) in a blue-shifted component at $\sim$0 km~s$^{-1}$, northward, and a red-shifted component at $\sim$15 km~s$^{-1}$, southward of the H{\sc ii} region. 
Also, the hydroxyl masers in ON\,1 provide some confirmation of this scenario \citep{namm:2006,fish:2007}. 
The expansion velocities of the masers, calculated by assuming a linear expansion
from the center of the H{\sc ii} region, would be 5.8\,km~s$^{-1}$ and 3.6\,km~s$^{-1}$ for the northern and southern maser groups, respectively.

\subsection{L\,1206}

For L\,1206, the trigonometric parallax distance,
$0.776^{+0.104}_{-0.083}$~kpc, is shorter than the kinematic distance; 1.0
kpc \citep[based on a hydroxyl maser, at $v_\mathrm{LSR}=-8.5~\mathrm{km~s^{-1}}$, using the
Galactic rotation curve of \citet{wouterloot:1989} with $R_0=8.5$~kpc, $\Theta_0=220$~km~s$^{-1}$,][]{macleod:1998}, 1.23 kpc
\citep[ammonia, $v_\mathrm{LSR}=-9.9~\mathrm{km~s^{-1}}$, using the Galactic
rotation curve of \citet{brand:1993}, with $R_0=8.5$~kpc, $\Theta_0=220$~km~s$^{-1}$][]{molinari:1996} and 1.4 kpc (6.7
GHz methanol, flat rotation curve, $R_0=8.5$~kpc, $\Theta_0=220$~km~s$^{-1}$). 
Our result places L\,1206 in the Local spur. Cep A, a nearby ($3\rlap{$.$}\,\degr8$\
separation, P.A. 126 \degr) star-forming region, is located at a distance of
$700\pm39$ pc, as determined by a parallax measurement of 12.2 GHz methanol masers \citep{moscadelli:2009}. The L\,1206 SFR
therefore seems to be in the same part of the Local arm as Cep A. 
L\,1206 is a dark cloud with an infrared source, IRAS 22272+6358A, which coincides
with the methanol maser emission and a 2.7\,mm dust continuum peak
\citep{beltran:2006}. Since neither 2\,cm nor 6\,cm radio emission is detected \citep{wilking:1989,McCutcheon:1991}, L\,1206 is thought to be in a young phase prior to the formation of an H{\sc ii} region.
\citet{beltran:2006} report large CO outflows, and put the systemic velocity of the ambient medium between [$-13.5$, $-8.5$] km s$^{-1}$, consistent with our systemic velocity range of
[$-13.3$, $-10.9$] km s$^{-1}$. 

\subsection{L\,1287}

The parallax for L\,1287 sets it at a distance of $0.929^{+0.034}_{-0.033}$\,kpc. 
This is close to the photometric distance of $\sim$850\,pc from \citep{yang:1991}, which placed L\,1287 in the Local
arm. However, the kinematic distance based on the methanol maser line would place
L\,1287 at 2\,kpc in the Perseus arm. 
 
Methanol masers in the dark cloud L\,1287 are located at the base of the bipolar CO
outflow \citep{yang:1991} originating in the main core of the cloud. The
infrared point source (IRAS 00338+6312) in the center of the core indicates
a proto-stellar object surrounded by cold, high density  $\mathrm{NH_3}$ (1,1)
gas
\citep{estalella:1993}.

\subsection{NGC\,281-W}
\begin{figure} 
\centering
\includegraphics[width=10cm,angle=-90]{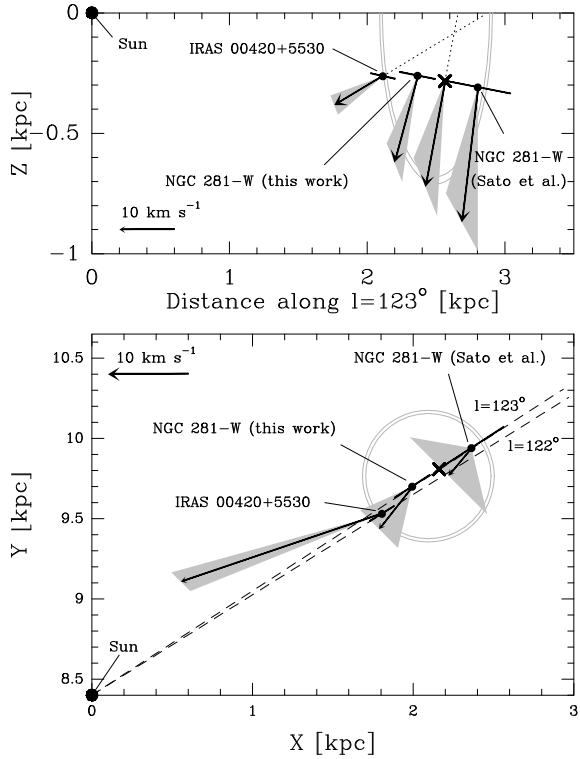}
\caption{Lower panel: a face-on view of
  the Galactic plane. The Galactic longitudes of NGC\,281-W ($l\simeq123\degr$) and IRAS
  00425530 ($l\simeq122\degr$) are indicated by dashed lines. The top panel
  shows a cross-section of the Galactic plane along the Galactic longitude of
  NGC\,281-W (123\degr). The vertical coordinate of this plot is the direction perpendicular
  to the Galactic plane. We plot NGC\,281-W and IRAS 00420+5530 as black
  dots, together with the respective distance error. Black arrows show the peculiar motion of the
  source after removing the Galactic rotation ($R_0=8.4$ kpc,
  $\Theta_0=254~\mathrm{km~s^{-1}}$). The errors in peculiar
  motion are indicated by the gray cones. The light-gray circles mark the super bubble with the center
of the bubble at a distance of 2.5 kpc and a radius of 0.4 kpc, adopted after
\citet{sato:2008}. The cross marks the averaged
  distance of 2.58~kpc. The dotted lines show the backward prolongation of the
  proper motion of NGC\,281-W and IRAS~00420+5530.}
\label{fig:bub}
\end{figure}

\label{sec:ngc}
There are numerous distance estimates for NGC\,281-W in the literature. Optical photometry places the cloud
between 2 and 3.68 kpc \citep{sharpless:1954, cruz-g:1974,  henning:1994,
guetter:1997}, while the kinematic distance is 3 kpc \citep{lee:2003}, based on a $v_\mathrm{LSR}=-30\,\mathrm{km~s^{-1}}$, using the Galactic rotation curve of \citet{clemens:1985} and $R_0=8.5\,\mathrm{kpc},\theta_0=220\,\mathrm{km~s^{-1}}$. The radial velocity of the methanol maser
line at $-29~\mathrm{km~s^{-1}}$, assuming a flat Galactic rotation curve, and $R_0=8.5\,\mathrm{kpc},\theta_0=220\,\mathrm{km~s^{-1}}$, would place NGC\,281-W at 2.5 kpc. 

Recently, \citet{sato:2008} reported a water maser parallax value of $0.355\pm0.030$\,mas, measured with the VERA
interferometer, corresponding to a distance of
$2.82\pm0.24$\,kpc. We measured a parallax of $0.421\pm0.022$\,mas, arriving at a
distance of $2.38^{+0.13}_{-0.12}$\,kpc. These distance measurements agree
with \citet{sato:2008} within 2$\sigma$ of the joint uncertainty.  Also, our proper 
motion agrees within 1$\sigma$ with the proper motion by \citet{sato:2008}, and
our radial velocity ($v_\mathrm{LSR}\sim-29$\,km~s$^{-1}$) is close to that of the
water masers ($v_\mathrm{LSR}\sim-31$\,km~s$^{-1}$). It is unlikely that the water
and methanol masers originate at such different depths into the SFR that it
would affect the distance. For example, in the SFR W3(OH), the water and
methanol maser parallaxes are in good agreement \citep{xu:2006,hachi:2006}.

NGC\,281-W is an SFR that lies near the edge of a super bubble in the Perseus
arm \citep{megeath:2003,Sato:2007}. Our slightly shorter distance would change 
the position of NGC\,281-W on this super bubble. Another SFR located on this
bubble has a parallax determined distance: IRAS 00420+5530 at
$2.17\pm0.05$~kpc \citep{moellen:2009}. This allows to compare the distances
and peculiar motions of both SFRs, and check whether the expanding super bubble, 
with the expansion center at 2.5~kpc adopted after
\citet{sato:2008}, can be responsible for the peculiar motions of both of them. 
We calculated the peculiar motions of IRAS 0042+5530 after \citet{moellen:2009} and  NGC 281-W after \citet{sato:2008} and our results. The bottom panel of Fig.~\ref{fig:bub} shows the peculiar motions in a face-on view of the Galactic plane. From the figure can be seen that the $U$ component of NGC\,281-W, representing the vector toward the Galactic center, is positive and non-zero in both our study and the work of \citet{sato:2008}. This means that NGC\,281-W is moving toward the Galactic center, which is the expected movement for the near side of an expanding bubble. In the top panel of Fig.~\ref{fig:bub} the peculiar motions of both SFRs are shown projected on a cross-section of the Galactic plane along the longitude of NGC 281-W, $l=123\degr$.
The average distance to NGC\,281-W based on \citet{sato:2008} and this work,
$2.58^{+0.27}_{-0.23}$~kpc, is marked in this panel as a cross. If we prolong the peculiar motion backwards of both IRAS 00420+5530 and the average result for NGC 281-W, they intersect at a distance from the Sun of 2.65~kpc and at a Galactic latitude of $\sim$$-0\rlap{$.$}\,\degr1$. 
This suggets that, if indeed both SFRs are expanding linearly from one expansion center, the expansion center is offset from the previously assumed position.
However, this offset center of expansion is difficult to understand when
looking at the peculiar motion of both SFRs in the Galactic plane
(Fig.~\ref{fig:bub}, bottom panel), since the motion does not seem to originate in one mutual expansion center; hence, it is difficult to pinpoint the characteristics of the super
bubble if the peculiar motions and distances of only two SFRs are all that are known.

\subsection{S\,255}
\label{sec:s255}
For S\,255, the parallax distance, $1.59^{+0.07}_{-0.06}$\,kpc, is much closer than the
commonly used photometric distance of 2.5\,kpc \citep{moffat:1979,blitz:1982}. S\,255 is an individual H{\sc ii}
region, associated with a complex of H{\sc ii} regions. The methanol maser emission
coincides with a filament of cold dust and molecular gas between two H{\sc ii} regions, S\,255
and S\,257. \cite{minier:2007} studied the star formation in this filament,
which they propose is possibly triggered by the compression of the filament by
the two H{\sc ii} regions. They find several molecular clumps in this
filament. At distance of $\approx2.5$\,kpc, the masses of the clumps, determined from the
submillimeter dust continuum, are around 300\,$\mathrm{M_\odot}$
\citep{minier:2007}. However, if we place S\,255 at the parallax determined distance of 1.6\,kpc, the clump masses would drop by 60\% to $\sim120$\,$\mathrm{M_\odot}$.

\section{Summary}

We measured parallaxes of 6.7 GHz methanol masers using the EVN towards five SFRs, achieving accuracies approaching 20~$\mathrm{\mu}$as. The primary results are summarized as follows.

\begin{itemize}
\item[1.]{We report trigonometric parallaxes for five star-forming regions, the distances to these
     sources are $2.57^{+0.34}_{-0.27}$~kpc for ON\,1, $0.776^{+0.104}_{-0.083}$ kpc for L\,1206, $0.929^{+0.034}_{-0.033}$ kpc for L\,1287, $2.38^{+0.13}_{-0.12}$ kpc for NGC\,281-W, and $1.59^{+0.07}_{-0.06}$ kpc for S\,255.}
\item[2.]{Galactic star-forming regions lag circular rotation on average by $\sim$17~km~s$^{-1}$, a
     value comparable to those found recently by similar studies \citep{reid:2009b}.}
\item[3.]{Individual 6.7 GHz methanol maser spots are stable over a period of
    $>2$\,years for most of the maser spots. The internal motions of the maser spots are weak ($\sim0.5-1$~km~s$^{-1}$) and rectilinear.}
\item[4.]{Measurements at 6.7 GHz are less disturbed by the troposphere, as expected. However, the ionospheric delay cannot be ignored and is likely not to be completely removed by using the JPL GPS--IONEX maps to calculate propagation delays. Continuum measurements of background sources show that most of them have significant structure that is evident from the large (up to 1--2 mas, or 2--10 km~s$^{-1}$) apparent movements between pairs of background sources. This additional uncertainty from the apparent movement does not prohibit a determination of the peculiar motion; however, it increases the error bars of the peculiar motion by the order of the uncertainty, depending on the distance and longitude of the source. The parallax uncertainty can be affected as well, if the cause of the apparent movement between the background sources are internal structure changes, which do not need to be perfectly linear. Two background sources should be a minimum for astrometric measurements at 6.7 GHz, so three or more background sources are recommended.}
\end{itemize}
\begin{acknowledgements}
The European VLBI Network is a joint facility of European, Chinese, South
African, and other radio astronomy institutes funded by their national research
councils. The National Radio Astronomy Observatory is a facility of the
National Science Foundation operated under cooperative agreement by Associated
Universities, Inc. We thank the staff at JIVE, especially Bob Campbell, for technical help and support.
We thank Amy Mioduszewski (NRAO) for her help with VLA frequency setup.
KLJR was supported for this research through a stipend from the
International Max Planck Research School (IMPRS) for Astronomy and
Astrophysics at the Universities of Bonn and Cologne. YX was supported by Chinese NSF through grants NSF 10673024, 10703010, and 10621303.  
\end{acknowledgements}

\bibliographystyle{aa}
\bibliography{13135bib}

\clearpage 
\onecolumn

\addtocounter{table}{1}

\longtab{5}{
\begin{longtable}{c c c c c}
\caption{\label{ta:data}Detailed results of parallax and proper motions measurements}\\
\hline
\hline
Background & $v_\mathrm{LSR}$ & Parallax & $\mu_\alpha$ & $\mu_\delta$\\
source     & ($\mathrm{km~s^{-1}}$)    & (mas)    & ($\mathrm{mas~yr^{-1}}$) & ($\mathrm{mas~yr^{-1}}$)\\
\noalign{\smallskip}
\hline
\noalign{\smallskip}
\endfirsthead
\caption{continued.}\\
\hline
\hline
Background & $v_\mathrm{LSR}$ & Parallax & $\mu_\alpha$ & $\mu_\delta$\\
source     & ($\mathrm{km~s^{-1}}$)    & (mas)    & ($\mathrm{mas~yr^{-1}}$) & ($\mathrm{mas~yr^{-1}}$)\\
\noalign{\smallskip}
\hline
\noalign{\smallskip}
\endhead
\multicolumn{5}{c}{{\bf ON\,1}}\\
\noalign{\smallskip}
\hline
\noalign{\smallskip}
\multicolumn{5}{c}{{\bf Northern group}}\\
J2003+3034 &-0.4      & $0.521\pm0.055~$ & $-3.33\pm0.18$ & $-5.26\pm0.41$\\ %c
           &\,0.0       & $0.401\pm0.052~$ & $-3.72\pm0.17$ & $-5.26\pm0.21$\\ %c
           &\,0.4       & $0.361\pm0.052~$ & $-3.82\pm0.17$ & $-5.28\pm0.19$\\  %c
           &\,0.7       & $0.526\pm0.057~$ & $-3.36\pm0.19$ & $-5.19\pm0.34$\\ %check
&\multicolumn{1}{l}{Combined fit} & $0.299\pm0.112^1$ &&\\
&\multicolumn{1}{l}{Averaging data} &$0.391\pm0.061~$ &&\\
\noalign{\smallskip}
J2009+3049 &-0.4    & $0.380\pm0.121~$ & $-2.87\pm0.39$  & $-4.88\pm0.31$\\ %c
           &\,0.0     & $0.459\pm0.140~$ &$-2.68\pm0.46$   & $-4.80\pm0.22$\\
           &\,0.4     & $0.483\pm0.142~$ &$-2.59\pm0.48$   & $-4.79\pm0.20$\\
           &\,0.7     & $0.415\pm0.112~$  & $-2.81\pm0.37$ & $-4.78\pm0.23$\\ %check
&\multicolumn{1}{c}{Combined fit} &$0.482\pm0.137^1$ &&\\
&\multicolumn{1}{c}{Averaging data} &$0.368\pm0.070~$ &&\\
\noalign{\smallskip}
\multicolumn{5}{c}{{\bf Southern group} }\\
J2003+3034 & 14.4     &   & $-3.88\pm0.12$ & $-5.92\pm0.14$\\
           & 14.8     &   & $-3.75\pm0.09$ &$-6.03\pm0.11$\\
           & 15.1     &   & $-3.67\pm0.21$ &$-6.09\pm0.19$\\
\noalign{\smallskip}
J2009+3034 & 14.4     &   & $-3.00\pm0.15$ & $-5.49\pm0.20$\\
           & 14.8     &   & $-2.88\pm0.21$ &$-5.59\pm0.26$\\
           & 15.1     &   & $-2.77\pm0.29$ &$-5.65\pm0.35$\\
\noalign{\smallskip}
\hline
\noalign{\smallskip}
Both QSOs & \multicolumn{1}{l}{Combined fit}   &$0.390\pm0.116\footnote{The error of
  the combined fit multiplied by $\sqrt{N}$, where $N$ is the number of maser spots.}$& & \\
          & \multicolumn{1}{l}{Averaging data} &$\mathbf{0.389\pm0.045~}$&&\\
$<\mu>_{north}(\sigma)_{pm}$ & &  & $-3.15\pm0.89(0.44)\footnote{We calculated
  an unweighted arithmetic mean of the individual proper motion results from
  all maser spots and background sources. The error bar on the mean is the
  standard error of the mean to which was added, in quadrature, the apparent movement between the two background sources of the respective coordinate. The uncertainty in the proper motion, which is introduced by the background source, is hereby taken into account. In parenthesis is given the standard deviation of the mean.}$ &$-5.03\pm0.46(0.22)\,^2$\\
$<\mu>_{south}(\sigma)_{pm}$ & &  & $-3.33\pm0.90(0.45)\,^{2}$ &$-5.80\pm0.46(0.23)\,^2$ \\
\noalign{\smallskip}
%\hline
\hline\hline
\noalign{\smallskip}
\multicolumn{5}{c}{{\bf L\,1206}}\\
\noalign{\smallskip}
\hline
\noalign{\smallskip}
J2223+6249 &-13.3     & $1.163\pm0.222~$ & $0.24\pm0.85$ & $-2.84\pm0.40$\\
           &-12.9     & $1.116\pm0.263~$ & $0.08\pm0.92$ & $-2.59\pm0.48$\\
           &-10.9     & $1.083\pm0.170~$ & $0.23\pm0.90$ & $-2.85\pm0.23$\\
           &-10.5     & $1.485\pm0.190~$ & $0.56\pm0.31$ & $-1.61\pm1.42$\\
           &\multicolumn{1}{l}{Combined fit} & $1.318\pm0.282^1$ & &\\
           &\multicolumn{1}{l}{Averaging data}&$1.331\pm0.180~$&&\\
\noalign{\smallskip}
J2225+6411 &-13.3           & $1.311\pm0.408~$ &$0.22\pm0.51$ & $-0.28\pm0.63$\\
           &-12.9           & $1.322\pm0.386~$ &$0.41\pm0.50$ & $-0.65\pm0.56$\\
           &-10.9           & $1.300\pm0.416~$ & $0.35\pm0.52$ & $-0.50\pm0.65$\\
           &-10.5           & $1.174\pm0.237~$ & $0.05\pm0.64$ & $~~\,0.11\pm0.27$\\
           &\multicolumn{1}{l}{Combined fit}& $1.272\pm0.384^1$&&\\
           &\multicolumn{1}{l}{Averaging data}&$1.288\pm0.241~$&&\\
\noalign{\smallskip}
\hline
\noalign{\smallskip}
Both QSOs  &\multicolumn{1}{l}{Combined fit}&$1.331\pm0.250^1$&&\\
           &\multicolumn{1}{l}{Averaging data}&$\mathbf{1.289\pm0.153}~$&&\\
$<\mu>(\sigma)_{pm}$ & & & $0.27\pm0.23(0.16)^2$ &$-1.40\pm1.95(1.15)\,^2$\\
\noalign{\smallskip}
\hline
\hline
\noalign{\smallskip}
\multicolumn{5}{c}{{\bf L\,1287}}\\
\noalign{\smallskip}
\hline
J0035+6130&-27.0          &$1.111\pm0.074~$ &$-0.18\pm0.10$ &$-2.30\pm0.25$\\
          &-23.9          &$0.928\pm0.078~$ &$-1.02\pm0.12$ &$-2.28\pm0.08$\\
          &-23.5          &$0.957\pm0.093~$ &$-0.93\pm0.11$ &$-2.48\pm0.12$\\
          &-23.2          &$1.002\pm0.084~$ &$-0.88\pm0.11$ &$-2.39\pm0.09$\\
          &-22.8          &$0.940\pm0.083~$ &$-1.14\pm0.11$ &$-2.78\pm0.10$\\
          &-22.5          &$0.917\pm0.067~$ &$-1.17\pm0.12$ &$-3.28\pm0.07$\\
          &\multicolumn{1}{l}{Combined fit}&$0.984\pm0.086^1$&&\\
          &\multicolumn{1}{l}{Averaging data} & $1.016\pm0.052~$&&\\
\noalign{\smallskip}
J0037+6236&-27.0           &$1.306\pm0.071~$ &$-0.14\pm0.10$ &$-1.71\pm0.40$\\
          &-23.9           &$1.244\pm0.044~$ &$-0.88\pm0.04$ &$-1.72\pm0.27$\\
          &-23.5           &$1.225\pm0.040~$ &$-0.81\pm0.04$ &$-1.77\pm0.25$\\
          &-23.2           &$1.011\pm0.101~$ &$-0.88\pm0.14$ &$-1.83\pm0.12$\\
          &-22.8           &$0.945\pm0.125~$ &$-1.17\pm0.20$ &$-2.20\pm0.14$ \\
          &-22.5           &$1.330\pm0.043~$&$-1.01\pm0.04$&$-2.64\pm0.37$\\
          &\multicolumn{1}{l}{Combined fit}&$1.192\pm0.107^1$&&\\
          &\multicolumn{1}{l}{Averaging data}& $1.150\pm0.052~$&&\\
\noalign{\smallskip}
\hline
\noalign{\smallskip}
Both QSOs &\multicolumn{1}{l}{Combined fit}&$1.079\pm0.069^1$&&\\
          &\multicolumn{1}{l}{Averaging data}& $\mathbf{1.077\pm0.039~}$&&\\
$<\mu>(\sigma)_{pm}$ & & &$-0.86\pm0.11(0.33)^2$& $-2.29\pm0.56(0.46)^2$\\
\noalign{\smallskip}
\hline
\hline
\noalign{\smallskip}
\multicolumn{5}{c}{{\bf NGC\,281-W}}\\
\noalign{\smallskip}
\hline
\noalign{\smallskip}
J0047+5657 &-30.2     &$0.416\pm0.054~$&$-2.64\pm0.15$ &$-1.75\pm0.10$\\
           &-29.9     &$0.380\pm0.045~$&$-2.54\pm0.12$ &$-1.72\pm0.08$\\
           &-29.5     &$0.401\pm0.041~$&$-2.58\pm0.19$ &$-1.72\pm0.07$\\
           &-29.2     &$0.404\pm0.047~$&$-2.69\pm0.22$ &$-1.75\pm0.08$\\
           &-28.8     &$0.497\pm0.026~$&$-2.55\pm0.04$ &$-1.69\pm0.13$\\
           &-28.1     &$0.529\pm0.028~$&$-2.68\pm0.04$ &$-1.52\pm0.14$\\
           &\multicolumn{1}{l}{Combined fit}&$0.400\pm0.070^1$ &&\\
          &\multicolumn{1}{l}{Averaging data}& $0.398\pm0.042~$ &&\\
\noalign{\smallskip}
J0052+5703 &-30.2      &$0.459\pm0.049~$&$-2.79\pm0.06$ &$-1.87\pm0.27$\\
           &-29.9      &$0.420\pm0.048~$&$-2.68\pm0.05$ &$-1.84\pm0.25$\\
           &-29.5      &$0.388\pm0.048~$&$-2.75\pm0.05$ &$-1.84\pm0.22$\\
           &-29.2      &$0.408\pm0.054`$&$-2.87\pm0.07$ &$-1.87\pm0.20$\\
           &-28.8      &$0.602\pm0.056~$&$-2.65\pm0.07$ &$-1.82\pm0.26$\\
           &-28.1      &$0.648\pm0.055~$&$-2.76\pm0.07$ &$-1.65\pm0.34$\\
           &\multicolumn{1}{l}{Combined fit}&$0.399\pm0.054^1$ &&\\
          &\multicolumn{1}{l}{Averaging data}& $0.425\pm0.024~$&&\\
\noalign{\smallskip}
Both QSOs &\multicolumn{1}{l}{Combined fit}&$0.412\pm0.045^1$&&\\
          &\multicolumn{1}{l}{Averaging data}& $\mathbf{0.421\pm0.022~}$&&\\
$<\mu>(\sigma)_{pm}$ & & &$-2.69\pm0.16(0.10)^2$&$-1.77\pm0.11(0.10)\,^2$\\
\noalign{\smallskip}
\hline\hline
\noalign{\smallskip}
\multicolumn{5}{c}{{\bf S\,255}}\\
\noalign{\smallskip}
\hline
\noalign{\smallskip}
J0613+1708 & 4.6     & $\mathbf{0.628\pm0.027}~$ & $-0.14\pm0.05$ & $-0.84\pm1.67$\\
$\mu_{pm}$  &                &                 & $-0.14\pm0.54^2$ & $-0.84\pm1.76^2$\\
\noalign{\smallskip}
\hline
\hline
\end{longtable}
}
\end{document}